\documentclass[fleqn,usenatbib]{mnras}

\usepackage{newtxtext,newtxmath}

\usepackage[T1]{fontenc}

\DeclareRobustCommand{\VAN}[3]{#2}
\let\VANthebibliography\thebibliography
\def\thebibliography{\DeclareRobustCommand{\VAN}[3]{##3}\VANthebibliography}

\usepackage{graphicx}	
\usepackage{amsmath}	
\usepackage{amssymb}	
\graphicspath{{./}{figures/}}
\usepackage[version=3]{mhchem}
\usepackage{gensymb}
\usepackage{multirow}

\providecommand{\e}[1]{\ensuremath{\times 10^{#1}}}

\newcommand{\fig}{Fig.}

\newcommand{\figref}[1]{\fig~\ref{#1}}

\newcommand{\tabref}[1]{table~\ref{#1}}
\newcommand{\Tabref}[1]{Table~\ref{#1}}

\renewcommand{\eqref}[1]{Eq.~(\ref{#1})}
\newcommand{\Eqref}[1]{Equation~(\ref{#1})}

\newcommand{\secref}[1]{Section~\ref{#1}}

\title[Adsorption \& Desorption of Nitrogen on ASW]{Neural-Network Assisted Study of Nitrogen Atom Dynamics on Amorphous Solid Water. I. Adsorption \& Desorption}

\author[Germ\'an Molpeceres et al]{
Germ\'an Molpeceres,$^{1}$\thanks{E-mail: molpeceres@theochem.uni-stuttgart.de (GM)}
Viktor Zaverkin,$^{1}$\thanks{E-mail: zaverkin@theochem.uni-stuttgart.de (VZ)}
Johannes K\"astner$^{1}$\thanks{E-mail: kaestner@theochem.uni-stuttgart.de (JK)}
\\
$^{1}$Institute for Theoretical Chemistry 
University of Stuttgart 
Pfaffenwaldring 55, 70569 
Stuttgart, Germany\\
}

\date{https://doi.org/10.1093/mnras/staa2891}

\pubyear{2020}

\begin{document}
\label{firstpage}
\pagerange{\pageref{firstpage}--\pageref{lastpage}}
\maketitle

\begin{abstract}
Dynamics of adsorption and desorption of ($^{4}$S)-N on amorphous solid water are analyzed using molecular dynamics simulations. The underlying potential energy surface was provided by machine-learned interatomic potentials.
Binding energies confirm the latest available theoretical and experimental results. The nitrogen sticking coefficient is close to unity at dust temperatures of $10$~K but decreases at higher temperatures. We estimate a desorption time scale of 1 $\mu$s at 28~K. The estimated time scale allows chemical processes mediated by diffusion to happen before desorption, even at higher temperatures.
We found that the energy dissipation process after a sticking event happens on the picosecond timescale at dust temperatures of $10$~K, even for high energies of the incoming adsorbate. Our approach allows the simulation of large systems for reasonable time scales at an affordable computational cost and \textit{ab-initio} accuracy. Moreover, it is generally applicable for the study of adsorption dynamics of interstellar radicals on dust surfaces.
\end{abstract}




\section{Introduction} \label{sec:intro}

Surface processes are of great importance for explaining molecular abundances in astrophysical environments \citep{Watanabe2008}. The formation of the simplest molecule, \ce{H2}, can be exclusively explained considering surface processes \citep{Cuppen2017}. However, surface chemistry in space is not limited to the archetypal \ce{H2} formation and the amount of possible chemical paths in surfaces increases constantly, determined both experimentally \citep{Kobayashi2017, Fedoseev2017, Qasim2019, Potapov2019, Oba2019, Potapov2020} and theoretically \citep{Enrique-Romero2016, Meisner2017, Rimola2018, Molpeceres2019, Lamberts2019}. The composition of solid bodies (known as dust grains) in the interstellar medium depends on the temperature conditions of each particular astronomical object and the nature of the parent star. However, at cryogenic temperatures in many cases, a refractory core of amorphous (hydrogenated or not) carbon or silicate is covered by amorphous ice, mainly composed of \ce{H2O} or \ce{CO} but with other components in lower abundance. 


Three physical processes are essential to describe surface chemistry in the interstellar medium (ISM). These are accretion, diffusion, and desorption. Accretion refers to the accumulation of molecules on top of the surface. Rates of accretion are modified by sticking coefficients, i.e., a fraction of molecules that remain bound to the surface after a collision. Diffusion and desorption refer to the mobility of the adsorbate on the surface and its return to the gas-phase media. We do not consider here irradiation processes (with, for example, photons, electrons, or cosmic-rays) \citep{Oberg2016}. Adsorbed molecules can either form a chemical bond with the surface (chemisorption) or bind weakly to it  (physisorption), allowing for surface diffusion. The solid effectively accommodates any excess energy from surface processes. Regardless of the nature of the bonding between adsorbate and surface, it is clear that dust grains play, if not a direct catalytic one, an important role in interstellar surface chemistry. 

Radical species, and specifically atoms, are a significant challenge for empirical force-fields and semi-empirical methods. Recently, the adsorption of atoms on water ice has been revisited \citep{Shimonishi2018} employing a cluster of 20 water molecules as a model for the ice. One main finding was that the binding energy of an adsorbate on a cluster of molecules can vary up to a factor of $\sim$ 3 from the binding energies calculated using a simplified dimer model \citep{Wakelam2017}. This is the case of the adsorption of ($^{4}$S)-N, which is predicted to have a very small binding energy of around 400--720~K \citep{Minissale2016,Shimonishi2018}. Such a change in binding energies has a deep impact on the chemistry of \ce{N} causing nitrogenation and formation of \ce{N2} to compete with hydrogenation (with formation of \ce{NH3}) \citep{Minissale2016,Shimonishi2018}, changing the chemical evolution of N-bearing species \citep{Linnartz2015}. These are key to the formation of complex organic molecules. Therefore, a precise description of adsorption and diffusion dynamics of the N atom is required to understand nitrogen chemistry.

Here, we report on simulations of the adsorption dynamics of nitrogen atoms in their $^{4}$S ground state on amorphous solid water using extensive sampling  on \textit{ab-initio} accuracy. The low binding energy of N on ASW allows us to sample a variety of different outcomes of the adsorption dynamics in reasonable computational timescales. To overcome the problem of lack of accuracy when dealing with radicals we use a variant of atomistic neural networks (ANN) \citep{Behler07}, recently developed in our group \citep{Zaverkin2020}. ANNs are used to construct a high quality potential that preserves the accuracy of DFT at a small fraction of its cost. ANN's popularity has grown in recent years in several research areas, but their application to problems in the astrochemistry is, to the best of our knowledge, new. Our intention with the present paper is to study sticking and desorption of N atoms on ASW, and to provide an atomistic insight into the topic for the astrochemical community.

\section{Methods} \label{sec:methods}

In order to be able to model the dynamics of a nitrogen atom with several hundred water molecules at sufficient accuracy, we employ ANNs in the form of a recently developed machine learning method, Gaussian moment neural networks (GM-NN) \citep{Zaverkin2020}, trained on DFT data.

Our methods are described as follows. First, we justify the choice of the functional and the construction of the training data set. Second, we shortly revise the machine learning method used in this work with a focus on its training. Third, we describe the use of the GM-NN potential in molecular dynamics (MD) simulations and the analysis of the obtained trajectories.

\subsection{Generation of GM-NN Training Data via AIMD Simulations} \label{sec:methods_data}

The construction of the data set used to train machine learning potentials (MLP), has to cover the space of relevant configurations of the system. To do so ab-initio molecular dynamics (AIMD) simulations were applied. Different levels of theory were used for \textbf{(1)} propagation and \textbf{(2)} refinement. The method used for \textbf{(1)} defines the potential energy surface on which structures are sampled. It influences the choice of the training geometries, but not the energies and forces on which the GM-NN method is trained. We employed the semiempiric methods GFN2-xTB \citep{Bannwarth2019} and PM7 \citep{Stewart2013} for propagation. The intention of \textbf{(1)} is to provide a quick exploration of a PES from which MD snapshots are extracted to be refined. An integration time step of 0.5~fs was employed in all simulations. No periodic boundary conditions are used in this work unless stated otherwise.

Refinement, \textbf{(2)}, is done at a higher level of theory and provides the energies and forces used for training. To find an appropriate DFT method, we benchmarked the performance of several exchange \& correlation functionals for the interactions within the N-\ce{H2O} system. The results can be found in Appendix \ref{sec:benchmark}. A very similar behavior for most of the tested DFT methods was observed. We decided to use the triple-corrected hybrid PBEh-3c \citep{Grimme2015}, with its associated basis set (def2-mSVP) for all refinement calculations. 

The training set can be divided into three different classes. The first class contains only water clusters and samples specific arrangements and interactions of water molecules in water ice to transfer them to MLPs. To create this class NVT simulations of water clusters of different sizes were performed. We have employed GFN2-xTB for \textbf{(1)} in these simulations. The water class data contains $1000$ snapshots of one water molecule at $800$~K, $1000$ snapshots of a water dimer at $300$~K, $1000$ snapshots of a water trimer at $300$~K, $800$ snapshots of a cluster of 30 water molecules at $150$~K, $400$ snapshots cluster of 30 water molecules at $800$~K, $200$ snapshots of a $60$ water cluster at $300$~K, $100$ snapshots of a $93$ water cluster $300$~K, $10$ snapshots of a $109$ water cluster at $300$~K, and $10$ snapshots of a $126$ water cluster at $300$~K. Note, that the high sampling temperatures are required to be able to reach high-energy regions of the potential. Those regions of configurational space with low energies are sampled most efficiently by molecular dynamics, rather independently of the temperature. Initial clusters were generated by random placement of water molecules and subsequent geometry optimization at the GFN2-xTB level of theory. The frequency of sampling of training points varied between $5$~fs for the small oligomers to $1$~ps for the big clusters. At high temperatures, a harmonic spherical boundary potential was applied during the sampling to ensure that all water molecules remain within a reasonable interaction limit. The parameters of the spherical potential have been chosen according to the maximum cluster size. The potential results in a restoring force acting towards the center of the cluster as soon as any atom is at 6.1~{\AA} from the geometric center of the cluster. The force constant of the potential was set to an arbitrarily high value of about 19 MJ/mol/\AA$^{2}$. Note, that the boundary potential was only applied in the sampling, not for the calculation of energies and forces for the training set.

\begin{table}
	\centering
	\caption{Structures included in the training data. Temperature, $T$, is given in K and $N$ is the total number of structures. The energy of the nitrogen atom in the case of collision dynamics sampling is given in parentheses in kJ/mol. Sampling interval for extracting refinement points, $f$, is given in fs.}
	\label{tab:training_data}
	\resizebox{\linewidth}{!}{\begin{tabular}{lccc|lccc} 
		\hline
		System      & $T$   & $N$  & $f$    & System                                & $T$   & $N$ &  $f$ \\
		\hline
		\ce{H2O}    & $800$ & $1000$ & 5  & 109 \ce{H2O}                           & $300$ & $10$  & 1000    \\
		2 \ce{H2O}   & $300$ & $1000$ & 5   & 126 \ce{H2O}                           & $300$ & $10$  & 1000   \\
		3 \ce{H2O}   & $300$ & $1000$ & 5  & \ce{N}-30 \ce{H2O}                     & $200$ & $500$  & 25  \\
		30 \ce{H2O}  & $150$ & $800$  & 12.5  & \ce{N}-7 \ce{H2O}                      & $150$ & $1000$  & 5 \\
		30 \ce{H2O}  & $800$ & $400$  & 25  & dyn. \ce{N}($2.4$)-7 \ce{H2O}           & $150$ & $17375$ & 12.5 \\
		60 \ce{H2O}  & $300$ & $200$  & 50  & dyn. \ce{N}($2.4$)-12 \ce{H2O}          & $150$ & $4018$  & 25 \\
		93 \ce{H2O}  & $300$ & $100$  & 100  & dyn. \ce{N}($48.2$)-12 \ce{H2O}         & $150$ & $1302$  & 25 \\
		\hline
	\end{tabular}}
\end{table}

The second class of training data aims to sample the interaction of the nitrogen atom adsorbed on the water surface. For this purpose, NVT simulations at two different temperatures of water clusters with an adsorbed nitrogen atom were performed. To propagate the trajectories, the PM7 semi-empirical Hamiltonian was used since we found that it reproduces the interaction potential of the N-\ce{H2O} system qualitatively well. For \textbf{(2)} PBEh-3c/def2-mSVP was used again. This second class of the training data contained $500$ snapshots of nitrogen at a $30$ water cluster at $200$~K and $1000$ snapshots of nitrogen at a $7$ water cluster $150$~K. We extracted snapshots every $25$~fs and $5$~fs, respectively.

The third class is dedicated to the description of the interaction of the nitrogen atom with the ice surface at medium and long distances of the nitrogen atom to the surface. This part of the training set is the most important one for sticking studies and, thus, it constitutes majority of training data. To sample the respective interactions NVE collision trajectories were run on two different clusters. The PM7 Hamiltonian was again used for \textbf{(1)}. The nitrogen atoms were propagated towards a pre-optimized water clusters with initial kinetic energies drawn from a Maxwell--Boltzmann velocity distribution at $150$~K. This third class of training data contains $17375$ snapshots from $70$ trajectories of a nitrogen atom with an initial kinetic energy of $2.4$ kJ/mol colliding with a $7$ water cluster, $4018$ snapshots from $26$ trajectories of a nitrogen atom with an initial kinetic energy of $2.4$ kJ/mol colliding with a $12$ water cluster, and $1302$ snapshots from $26$ trajectories of a nitrogen atom with an initial kinetic energy of $48.2$ kJ/mol colliding with a $12$ water cluster. We extracted a snapshot every $12.5$~fs for the cluster with seven water molecules and every $25$~fs for the other two. The training data was extracted only if the nitrogen atom was within the cutoff radius of 5.5~{\AA} of any water molecule, selected for this work according to the relevant interaction range, see below. A list of conditions  included in the training can be found in \Tabref{tab:training_data}.

With this reasonably big and heterogeneous training set, we sample the relevant interactions during an atom-surface collision event, which has to be introduced in the MLP. Additionally, sampling high temperatures of the clusters and high incoming energies of the nitrogen atom introduces situations to the MLPs that should be avoided, such as destruction of hydrogen bonds. This is done to reduce the extrapolation of the MLPs applied in this work. The completeness of the training set was tested by a Query-by-Committee-like approach, see below.

The codes we have employed for the generation of the training set are, respectively: a) GFN2-xTB calculations with the xTB code \citep{Bannwarth2019, Grimme2017} and DL-Poly/Chemshell as MD driver \citep{Sherwood2003, dlpoly, Chemshell} b) PM7 calculations with Gaussian16 \citep{Stewart2013, g16} and DL-Poly/Chemshell as MD Driver, c) DFT energy and gradient calculations with Turbomole 7.4.1 \citep{Grimme2015, Turbomole}.

\subsection{Construction and Training of the GM-NN Model} \label{sec:gmnn_model}

Among the suite of existing machine learning models, the Gaussian moment neural network (GM-NN) approach, \citep{Zaverkin2020} uses NNs to represent the high-dimensional potential energy surface (PES). A single potential energy $E$ of a molecular or solid-state structure is written as a sum of atomic energy contributions, $E = \sum_i E_i$. The `atomic' energy depends on the local environment of the respective atom within a predefined cutoff sphere of the radius $R_\text{c}$. It was set to $R_\text{c}=5.5$~{\AA} to account for electrostatic interactions between the adsorbate (N atom) and the water ice surface. The description of the local environment is given in the GM-NN model through a set of novel symmetry-preserving local atomic descriptors, i.e. Gaussian Moments (GM). In addition to the geometric information, GMs include information about the atomic species of both the central and neighbor atoms. Therefore, for all atomic energy contributions, only a single NN has to be trained, in contrast to using an individual NN for each species as frequently done in the literature. The computational cost and memory usage of the GM-NN model scale linearly with the system size because  atomic neighbor lists are employed. Therefore, GM-NN is suitable for the applications on large systems like the ones studied in this work. We use a shallow neural network with two hidden layers consisting of $256$ and $128$ nodes (abbreviated GM-sNN in \citep{Zaverkin2020}).

To train the GM-sNN model the loss function 
\begin{equation}
    \mathcal{L} = w_\text{E}\left|\left| \Delta E\right|\right|^2 + \frac{w_\text{F}}{3N_\text{at}}\sum_{i=1}^{N_\text{at}}\left|\left| \Delta \mathbf{F}_i\right|\right|^2
\end{equation}
is minimized. Here, $N_\text{at}$ is the number of atoms in the respective structure, $\Delta E$ and $\Delta \mathbf{F}_i$ are the differences between the GM-sNN prediction and the reference data for energies and forces, respectively. The parameters $w_\text{E}$ and $w_\text{F}$ were set to $1~\text{au}$ and $100$~au~{\AA}$^2$, respectively. The network was trained using the AMSGrad optimizer \citep{Reddi18} with $32$ molecules per mini-batch. The learning rate was set to $10^{-3}$ and kept constant throughout the whole training procedure \citep{Zaverkin2020}.
The total data set constructed in \secref{sec:methods_data} was split into two subsets. About $86$~\% of the data was used to train the GM-sNN models, and the remaining $14$~\% were used for validation.

To verify the generalization quality of the constructed GM-sNN PES, we utilize a Query-by-Committee-like approach, frequently used in the active learning literature \citep{settles.tr09}. For this purpose, three models were trained using the same training data but with different randomly initialized NN parameters. During the dynamics,  Every $100$~fs, we evaluated the energy and atomic forces of the entire system using an ensemble of trained models. As long as the deviation between their predication is small, the structure is well-represented within the training set and the GM-NN PES generalizes well.  
For more details on the approach applied in this work, see Appendix \ref{sec:extrapolation}.  

Training of all models was performed within the Tensorflow framework \citep{TF15} for 5000 epochs on an NVIDIA GeForce GTX-1080-Ti-11GB GPU each. The training of the GM-NN model required at most four days.  

To run MD simulations with the machine learned potentials (MLP), the GM-NN approach was interfaced to the ASE package (v. 3.19.0) \citep{HjorthLarsen2017}.
 
\subsection{Simulations and Analysis} \label{sec:simulation}

\subsubsection{Initial Surface Construction}

The first, non-optimized surface is constructed in a similar way as in previous works of our group \citep{Molpeceres2020}. First, a sufficiently large cell of pre-equilibrated liquid TIP3P \citep{Jorgensen1981} water was generated using the solvate plugin within VMD v.1.9.1 \citep{VMD} with a density of $\sim$ 1.0~g~cm$^{-3}$. The cell comprises 18937 water molecules and has dimensions of $80\times 80 \times 80$~\AA. Periodic boundary conditions were applied in all directions. After an initial minimization, this liquid system was equilibrated for $100$~ps at $300$~K and then suddenly quenched to $10$ K for another $50$~ps. All these simulations were carried out in the NVT ensemble using a Langevin thermostat, using the NAMD code \citep{Phillips2005}.

The ice surface is modeled as a hemispherical cut of the quenched bulk ice. For this particular application, we have extracted a hemispherical cluster with a radius of $18$~\AA, encompassing 499 water molecules. The center of the hemisphere is set as the origin of coordinates for further calculations. A total of 295 water molecules are left active during the optimization and the dynamics, whereas the remaining 204 remain fixed to provide a boundary to the hemispherical cluster. 

\subsubsection{MLP-Based Simulations}
\begin{figure}
    \centering
        \includegraphics[width=7cm]{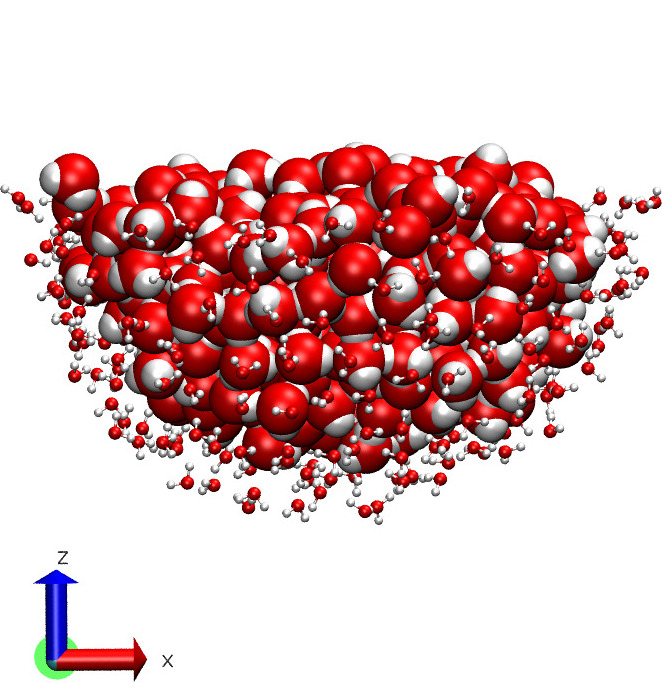}\\
        \includegraphics[width=7cm]{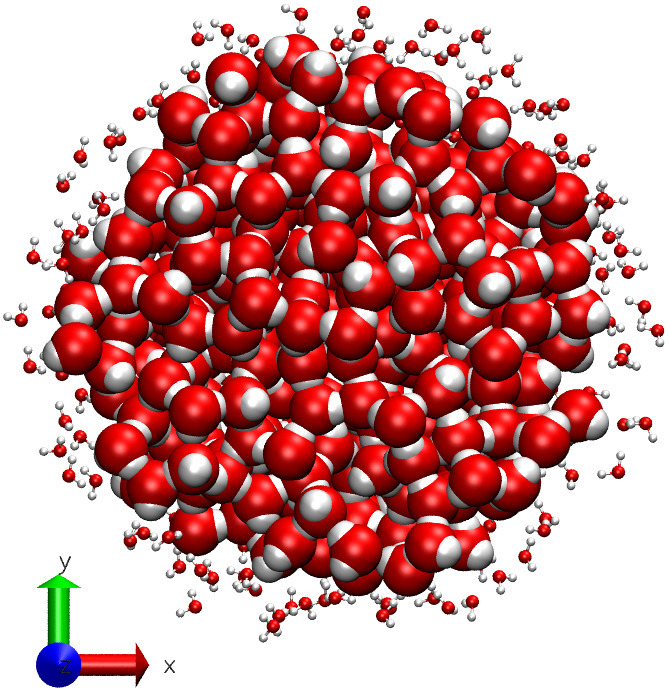}
    \caption{Side and top views of the water hemisphere used as a surface model for the amorphous water surface in our collision dynamics. Active atoms are represented as large spheres, frozen atoms at the boundary as balls and sticks.}
    \label{fig:asw_optimized}
\end{figure}

The geometry of the hemisphere is optimized using the LBFGS algorithm in the atomic simulation environment (ASE) before performing any MD runs. The optimized hemisphere is shown in \figref{fig:asw_optimized}.  We showed in our previous work that the process of sticking is virtually the same for internally flexible and internally rigid water models \citep{Molpeceres2020}. Here, we used flexible waters for all simulations.

The ice hemisphere was heated to 300 K for 255~ps to a disordered liquid state with a time step of 0.5~fs in an NVT ensemble using random initial velocities. After the first 10~ps a snapshot was taken every 5~ps to generate 49 independent ice surface structures. Subsequently, all 49 generated structures were quenched to 10~K, 50~K, 90~K, and 130~K for another 10~ps each. The temperature was controlled by a Langevin thermostat. The selection of temperatures for the ices was made to include very cold conditions (10--50 K) 
inside molecular clouds \citep{Snow2006} and warmer conditions that can be found in the outer layers of protoplanetary disks (50--150 K) \citep{Boss1998}. The upper bound of 130 K has been chosen as a boundary value before the known amorphous $\rightarrow$ crystalline phase transition, happening in water ice under astrophysical conditions \citep{Jenniskens1994}.

A similar approach to the one used in our previous work \citep{Molpeceres2020} was employed to sample collision dynamics. The collision point was randomly displaced laterally by up to 8~\AA{} from the center of the hemisphere. This, combined with different starting ice-snapshots ensures that every trajectory collides at a different ice configuration. Thus, the situation expected for amorphous ice is reproduced.

Given the collision point, the initial velocity vector from the adsorbate can be calculated. The polar ($\theta$) angles are sampled equidistantly, and the azimuthal ($\phi$) angles randomly. The theoretical maximal value for $\theta$ defined with respect to the surface normal is $90^{\circ}$.  In practice, an upper limit for $\theta$ is needed. The azimuthal angle was sampled between $0^{\circ}$ and $360^{\circ}$. We have used 15 different values for $\theta$ and 12 values for $\phi$. In the present work, translational energies ranging from 0.29~kJ/mol to 14.47~kJ/mol were chosen.

The collision trajectory is obtained by integrating classical equations of motion for the nuclei with the velocity-Verlet algorithm and a time step of 0.5~fs without any thermostat, i.e. in an NVE ensemble. Each trajectory is run until a bouncing event is detected or the maximum simulation time of 10~ps is reached. For each trajectory, the analysis has to differentiate between sticking and bouncing trajectories. 
Bouncing events were identified by analyzing the direction of the velocity of the adsorbate on the fly every 12.5~fs. In the case the velocity vector changed by $> 90^{\circ}$ to the initial velocity vector, the minimal distance to every atom on the ice surface is calculated. 
A trajectory is identified as a bounce if the minimal distance exceeds {6~\AA}. This condition was implemented as a stopping flag in our simulations. However, bouncing can also occur at lower angles. To cover these cases, we calculated the minimum distance to the ice hemisphere after 10~ps. If this was bigger than {6~\AA}, the trajectory was identified as bouncing. Any other trajectories are considered as a sticking event. Visual inspection of several trajectories confirmed the validity of these criteria.

The ratio of sticking vs. the total number of trajectories is the sticking probability $P\left(E\right)$. Bouncing and sticking events follow a binomial distribution, which we use to provide the statistical uncertainty using the Jeffreys interval, as implemented in the Statsmodels mathematical library \citep{Statmodels}. For a run with 15 values for $\theta$ and 12 values for $\phi$, we run a total of 180 trajectories per incoming energy, with a total of 9 incoming energies per surface temperature. 

\section{Results}\label{sec:results}

\subsection{Binding Energies of N on ASW} \label{sec:binding}

\begin{figure}
    \centering
    \includegraphics[width=\linewidth]{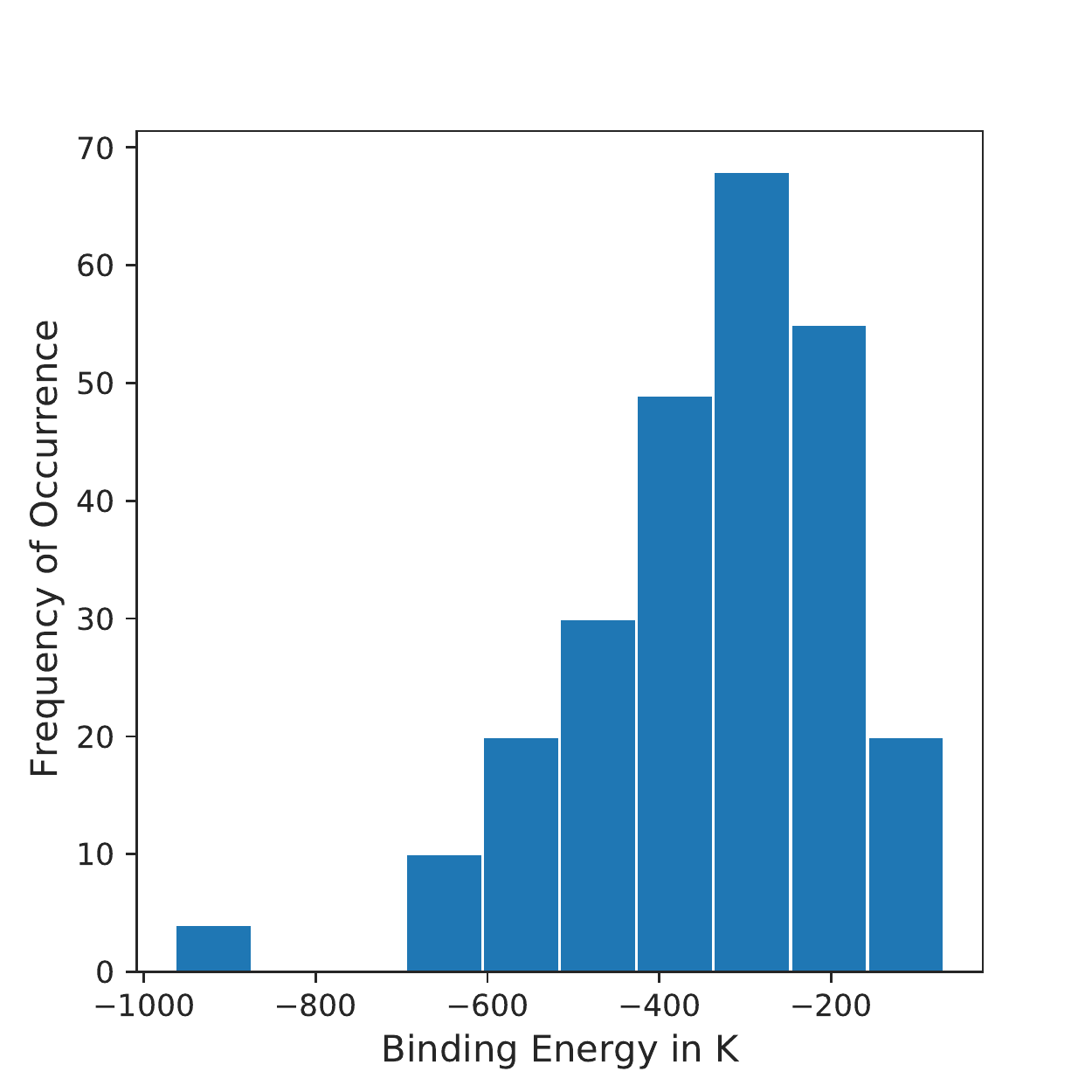}
    \caption{Distribution of binding energies for N atoms on ASW.}
    \label{fig:binding_energy}
\end{figure}

We calculated binding energies for several points on the ice surface by sampling a quadratic grid of 256 equidistant points with a step of $1$~{\AA}. The adsorbate was placed $3$~{\AA} above the surface, followed by a geometry optimization using our MLP. \figref{fig:binding_energy} shows the distribution of binding energies of the N-ASW system. The presented values are potential energy differences including zero-point vibrational energies (ZPE) for each configuration. The ZPE for each configuration was calculated using the analytic Hessian of the MLP. We obtained a mean of $-345.6$~K, where negative energies indicate exothermic binding, and a standard deviation of $155.0$~K. For comparison, without ZPE we obtained a mean of $-406.6$~K and the standard deviation is equal to $165.8$~K. During MD simulations, ZPEs are not taken into account and the latter mean value serves as an estimate of the average potential experienced by the adsorbate. From a chemical point of view, such values of the standard deviation are meaningful. It implies a broad distribution of binding sites as expected for an amorphous surface. Note that our  results for the binding energy are in excellent agreement with those presented by \cite{Shimonishi2018}, who report an average value of $400$~K, and close to the experimental value found by \cite{Minissale2016}, differing only by a factor of about 2. It is important to keep in mind that the distribution of binding energies is broad, implying that the factor is only an averaged value. This agreement of the binding energies with previous studies provides a solid justification for the applicability of the employed MLPs.

\subsection{Sticking Probabilities and Sticking Coefficients} \label{sec:sticking}

To study the sticking dynamics, we propagated a nitrogen atom with different incoming energies onto ice surfaces with different temperatures. We used the incoming energies $E= 0.29$, 0.48, 1.45, 2.41, 4.82, 7.24, 9.65, 12.06, and 14.47 kJ/mol and ice surfaces equilibrated at $T=10$~K, 50~K, 90~K, and 130~K. According to the criteria presented in \secref{sec:methods}, we identify trajectories as sticking or a bouncing events. After sampling, we determine the energy-dependent sticking probability, $P\left(E\right)$, as the ratio of sticking trajectories over the total number of trajectories.

\begin{table}
	\centering
	\caption{Sticking probabilities $P(E)$ and their 95\% confidence intervals as a function of the incoming energy of the adsorbate $E$ and the surface temperature $T_\text{s}$.}
	\label{tab:sticking_probabilities}
    \begin{tabular}{cccc}
    \hline
    \multicolumn{2}{c}{$T_\text{s}=10$~K} & \multicolumn{2}{c}{$T_\text{s}=50$~K}\\
    \hline
    $E$ (kJ/mol) & $P(E)$ & $E$ (kJ/mol) & $P(E)$  \\
      0.29     & 1.00 (0.99--1.00) &   0.29 & 0.99 (0.96--1.00) \\
      0.48     & 1.00 (0.99--1.00) &   0.48 & 0.98 (0.95--1.00) \\
      1.45     & 1.00 (0.99--1.00) &   1.45 & 0.99 (0.96--1.00) \\
      2.41     & 0.98 (0.96--1.00) &   2.41 & 0.98 (0.95--1.00) \\
      4.82     & 0.99 (0.96--1.00) &   4.82 & 0.94 (0.92--0.98) \\
      7.24     & 0.93 (0.88--0.96) &   7.24 & 0.93 (0.88--0.96) \\
      9.65     & 0.94 (0.90--0.97) &   9.65 & 0.86 (0.80--0.90) \\
     12.06     & 0.91 (0.86--0.94) &  12.06 & 0.78 (0.72--0.84) \\
     14.47     & 0.84 (0.78--0.89) &  14.47 & 0.78 (0.72--0.84) \\
     \hline
     \multicolumn{2}{c}{$T_\text{s}=90$~K} & \multicolumn{2}{c}{$T_\text{s}=130$~K}\\
    \hline
      0.29     & 0.93 (0.89--0.96) &   0.29 & 0.79 (0.72--0.84) \\
      0.48     & 0.95 (0.91--0.97) &   0.48 & 0.79 (0.73--0.85) \\
      1.45     & 0.95 (0.91--0.97) &   1.45 & 0.80 (0.74--0.85) \\
      2.41     & 0.91 (0.86--0.95) &   2.41 & 0.78 (0.72--0.84) \\
      4.82     & 0.86 (0.80--0.91) &   4.82 & 0.73 (0.66--0.79) \\
      7.24     & 0.81 (0.75--0.86) &   7.24 & 0.64 (0.56--0.71) \\
      9.65     & 0.73 (0.66--0.79) &   9.65 & 0.59 (0.52--0.66) \\
     12.06     & 0.64 (0.57--0.71) &  12.06 & 0.58 (0.51--0.65) \\
     14.47     & 0.63 (0.56--0.70) &  14.47 & 0.55 (0.48--0.62) \\
    \hline
    \end{tabular}
\end{table}

\begin{figure}
    \centering
    \includegraphics[width=\linewidth]{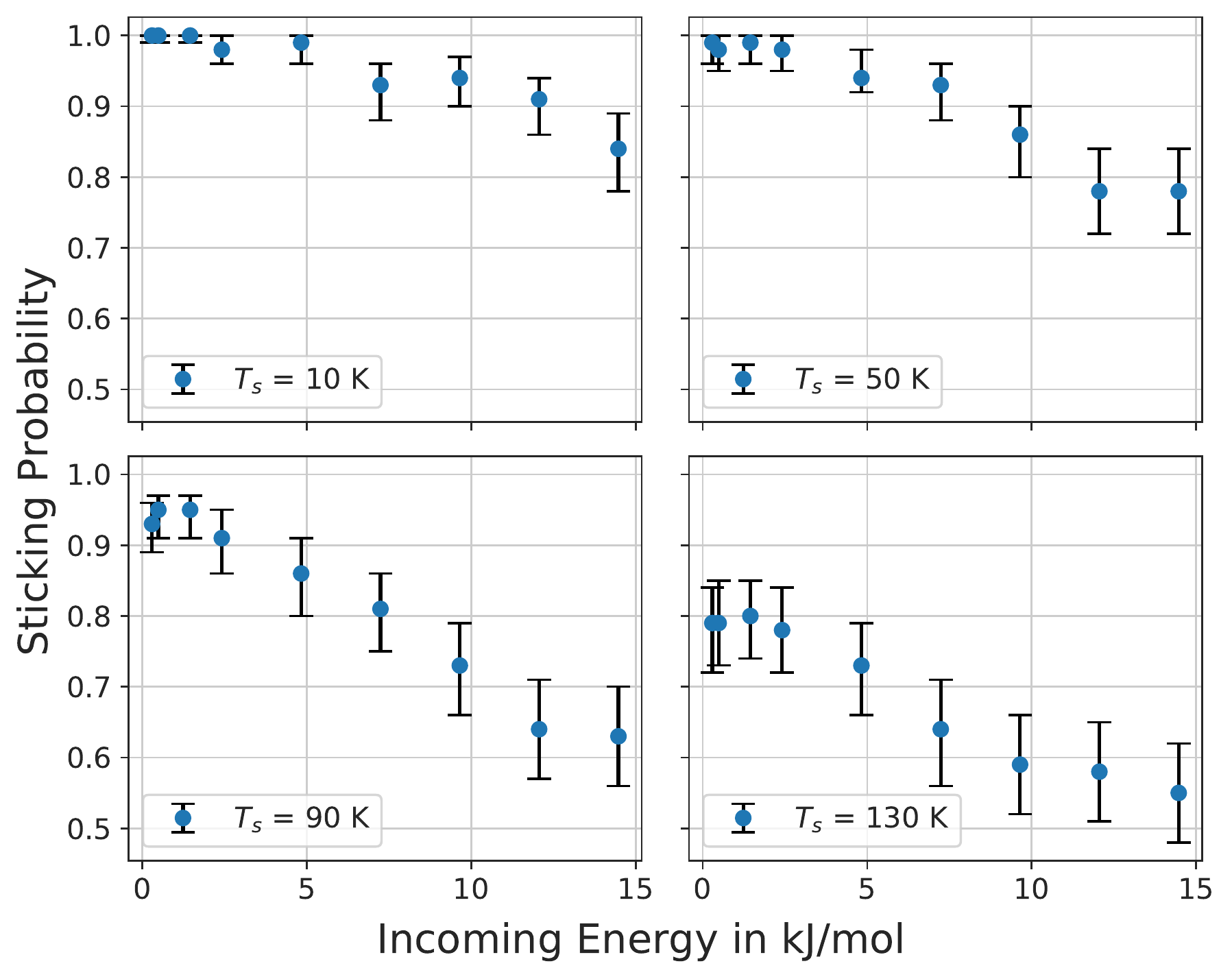}
    \caption{Sticking probabilities $P(E)$ at different adsorbate incoming energies and surface temperatures. Error bars correspond to a 95\% confidence interval.}
    \label{fig:sticking_probabilities}
\end{figure}

The obtained sticking probabilities are presented in \figref{fig:sticking_probabilities} and the respective numerical values are given in \Tabref{tab:sticking_probabilities}. 
At low $E$, we found a plateau (similar to the results in \cite{Molpeceres2020}) of predominantly sticking events. This plateau exists for all surface temperatures up to an incoming energy of 2.41 kJ/mol. Between $E=5$ and $10$~kJ/mol, a constant decrease of $P(E)$ with $E$ is observed. For higher incoming energies (above $10$~kJ/mol), this decrease is attenuated. 

Thermal sticking coefficients ($S_{T}$) are calculated by integrating sticking probabilities according to the expression
\begin{equation} \label{eq:integration}
    S_T = \frac{1}{(kT)^2}\int_0^\infty P\left(E\right)Ee^{-\frac{E}{kT}}\mathrm{d}E,
\end{equation}
with $k$ being the Boltzmann constant and $T$ the gas temperature. Such an integration has been employed previously either by approximating $P(E)$ with an exponential decay function \citep{Buch1991, Masuda1998} or by interpolation of the values for $P(E)$ obtained from the simulation and subsequent integration \citep{Molpeceres2020}. We have employed the latter approximation with $P(0)=1$ and $P(\infty)=0$. Error propagation from the sampling error in $P(E)$ allows an estimate of the statistical error in $S_T$. Details of its derivation are given in Appendix~\ref{sec:derivation}.

\begin{figure}
    \centering
    \includegraphics[width=\linewidth]{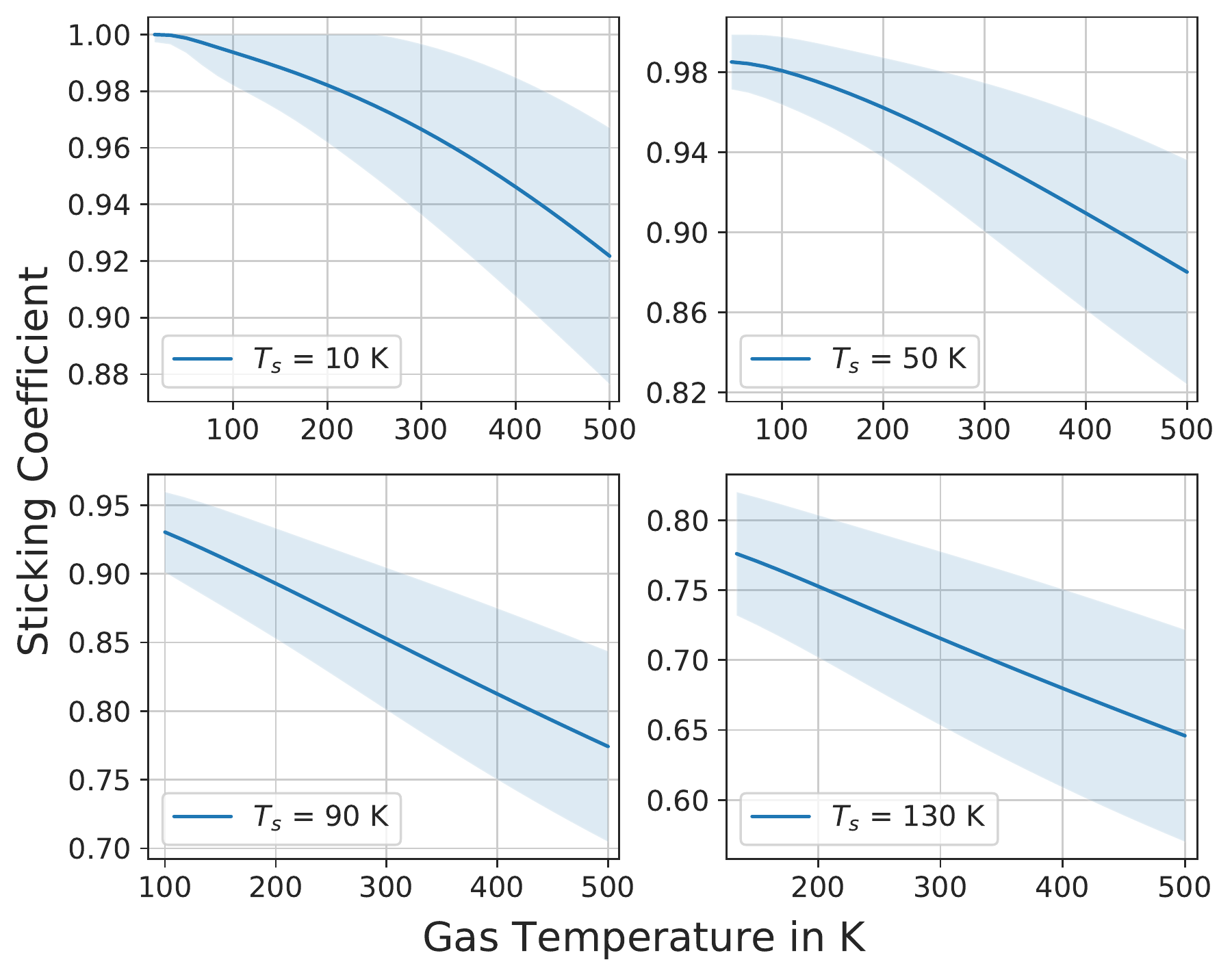}
    \caption{Thermal sticking coefficients $S_T$ at different surface temperatures. $95$~\% confidence intervals are obtained from error propagation. For all graphs the curve starts at $T_\text{g} = T_\text{s}$.}
    \label{fig:sticking_coefficients}
\end{figure}

\begin{table}
	\centering
	\caption{Sticking coefficients ($S_{T}$) as a function of the gas temperature ($T_\text{g}$) and the surface temperature.}
	\label{tab:sticking_coefficients}
    \begin{tabular}{cccc}
    \hline
    \multicolumn{2}{c}{$T_\text{s}=10$~K} & \multicolumn{2}{c}{$T_\text{s}=50$~K}\\
    \hline
      $T_\text{g}$ (K) & $S_{T}$ & $T_\text{g}$ (K) & $S_{T}$  \\
      10     & 1.00 (1.00--1.00) &   10 & 0.99 (0.98--1.00) \\
      33     & 1.00 (1.00--1.00) &   33 & 0.99 (0.97--1.00) \\
      50     & 1.00 (0.99--1.00) &   50 & 0.99 (0.97--1.00) \\
      83     & 1.00 (0.99--1.00) &   83 & 0.98 (0.97--1.00) \\
     100     & 0.99 (0.98--1.00) &  100 & 0.98 (0.96--1.00) \\
     150     & 0.98 (0.97--1.00) &  150 & 0.97 (0.95--0.99) \\
     250     & 0.97 (0.95--1.00) &  250 & 0.95 (0.92--0.98) \\
     300     & 0.96 (0.94--1.00) &  300 & 0.93 (0.90--0.98) \\
     450     & 0.93 (0.89--0.98) &  450 & 0.89 (0.84--0.95) \\
     \hline
     \multicolumn{2}{c}{$T_\text{s}=90$~K} & \multicolumn{2}{c}{$T_\text{s}=130$~K}\\
    \hline
      50     & 0.94 (0.92--0.96) &   50 & 0.79 (0.76--0.83) \\
      83     & 0.94 (0.91--0.96) &   83 & 0.79 (0.74--0.83) \\
     100     & 0.93 (0.90--0.96) &  100 & 0.79 (0.75--0.83) \\
     150     & 0.91 (0.88--0.95) &  150 & 0.77 (0.73--0.82) \\
     250     & 0.87 (0.83--0.92) &  250 & 0.73 (0.68--0.79) \\
     300     & 0.85 (0.80--0.90) &  300 & 0.72 (0.66--0.78) \\
     450     & 0.79 (0.73--0.86) &  450 & 0.66 (0.59--0.74) \\
    \hline
    \end{tabular}
\end{table}

The integrated sticking coefficients are listed in \Tabref{tab:sticking_coefficients} and illustrated in \figref{fig:sticking_coefficients}. From both the table and the figure, we observe that the sticking coefficients are close to unity for a surface temperature of 10~K and low gas temperatures. With the increase of the gas temperature, the sticking coefficients start to decrease. At a surface temperature $10$~K, the sticking coefficient drops by a total of $7\%$ at higher gas temperatures. Such a decrease can be significant in the case of thermally excited atoms coming from exothermic ion-molecule reactions. At higher surface temperatures, the trend starts to change, and the drop in the sticking coefficients varies between 10 and 16~\%. Starting at surface temperatures of $50$~K, the sticking coefficient is lower than unity for low gas temperatures. The finding of a sticking coefficient higher than zero at temperatures that we assume higher than the average desorption temperature is an interesting fact that requires further investigation in the next sections.

\subsection{Estimation of Desorption Times \& Temperatures}

Desorption of a species from a surface is an activated, uni-molecular process that can be described in the context of transition state theory by a classical Arrhenius equation
\begin{equation} \label{eq:desorption}
    k_\text{d}=\nu e^{\frac{- E_\text{bin}}{kT}}.
\end{equation}
Here, $k_\text{d}$ represents the rate constant for desorption, $\nu$ is the pre-exponential factor or attempt frequency, a measure of the frequency of the vibrational mode leading to desorption, and $E_\text{bin}$ is the (non ZPE) binding energy. 

\begin{figure}
    \centering
    \includegraphics[width=\linewidth]{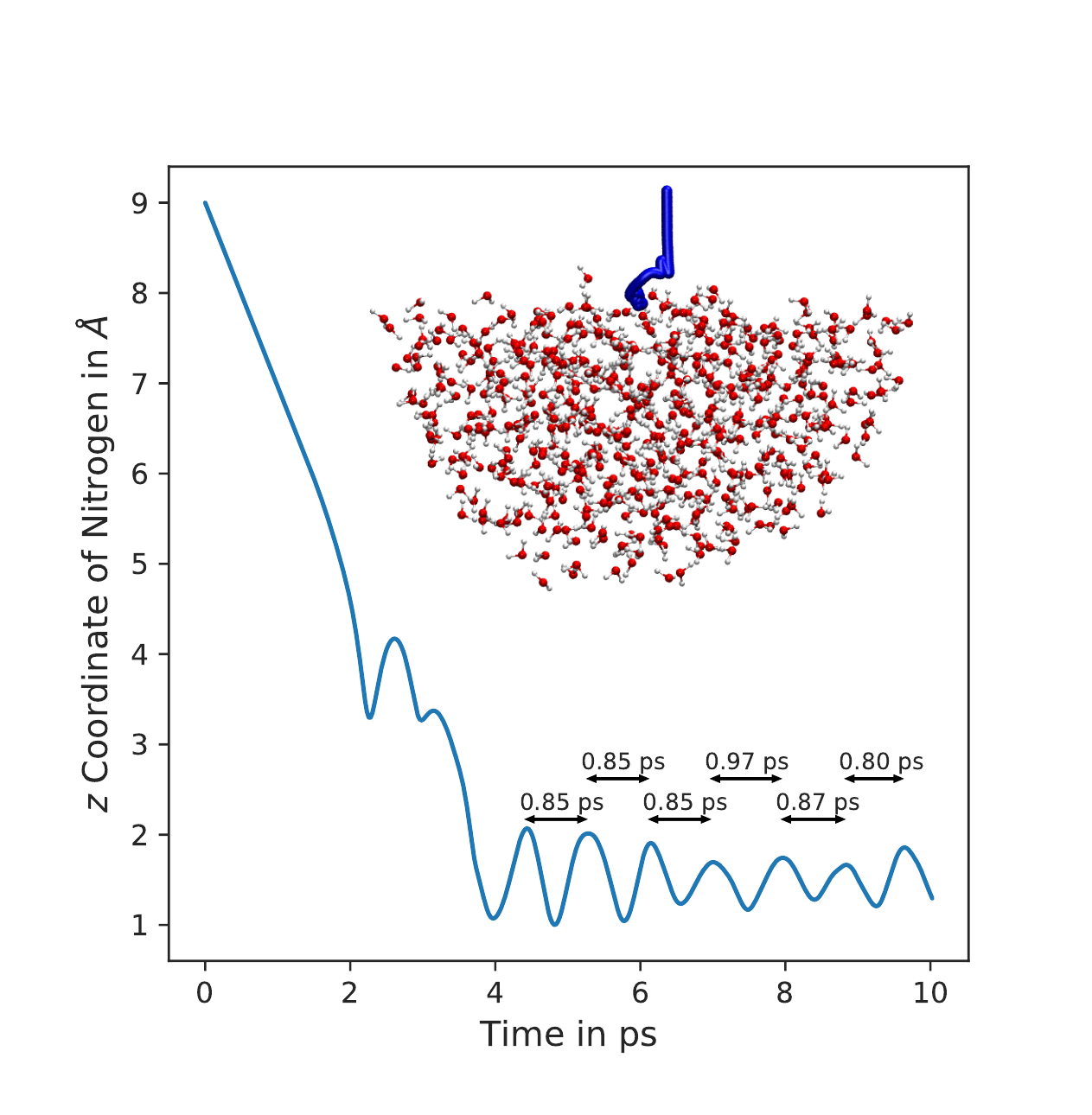}
    \caption{Time-dependence of the $z$ component of the nitrogen atom for an orthogonal collision trajectory ($T_\text{s}=10$~K, $E= 0.29$~kJ/mol, $\theta=0$\degree).}
    \label{fig:z_coord}
\end{figure}

The estimation of $E_\text{bin}$ has been presented in \secref{sec:binding}. The attempt frequency $\nu$ is, thus, the only quantity that remains unknown for the direct application of \Eqref{eq:desorption}. In good approximation, the vibrational mode leading to desorption from a surface will be dominated by movement orthogonal to the surface, i.e. in $z$ direction, which can directly be obtained from our molecular dynamics trajectories. We analyzed several trajectories and found good agreement between them. For illustration purposes, we show the variation of the $z$ component of the nitrogen atom in \figref{fig:z_coord} for one trajectory. For this image, a low incoming frequency of $E= 0.29$~kJ/mol a low surface temperature $T_\text{s}=10$~K and a collision direction orthogonal to the surface have been picked. 
Unmistakably, sticking can be observed after about $4$~ps. Vibrational periods are indicated in \figref{fig:z_coord}. Our mean value for the vibrational period of the nitrogen atom in $z$ direction is 0.86~ps, resulting in an attempt frequency of  $\nu=1/0.86=1.16$~ps$^{-1}$. With this value, rate constants and half-lives $\tau$ can be calculated: 
\begin{equation} \label{eq:desorption_halflives}
    \tau=\dfrac{\ln{2}}{k_\text{d}}.
\end{equation}

The results for this analysis are given in \Tabref{tab:half-lives}. The estimated half-lives are always above the time scales reachable in our simulations, which is why hardly any thermal desorption has been observed in our data. 
The important implication of these results is that it establishes an upper time limit to diffusion before thermal desorption.  

\begin{figure}
    \centering
    \includegraphics[width=\linewidth]{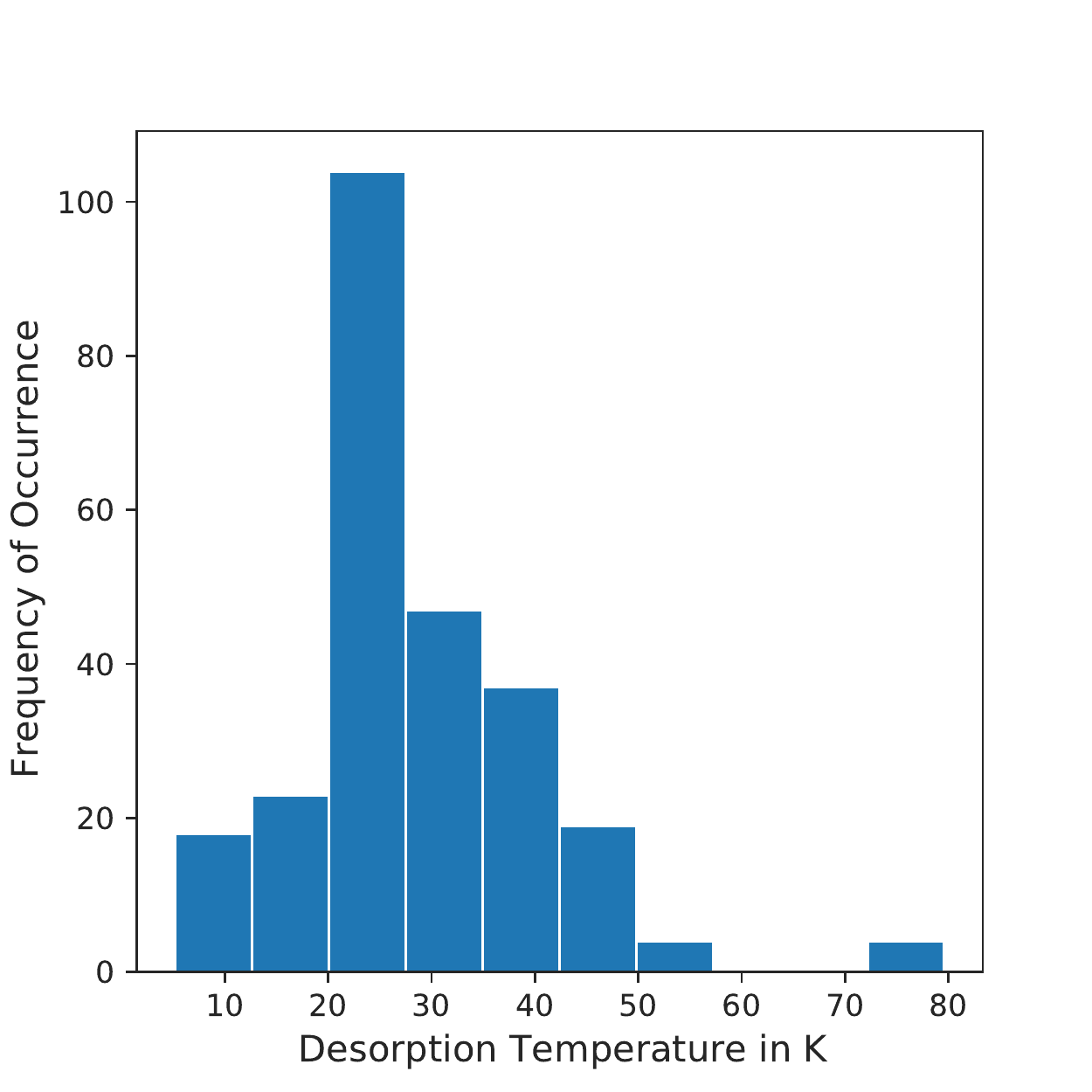}
    \caption{Distribution of desorption temperatures for N atoms on ASW assuming a desorption time-scale of 1~$\mu$s.}
    \label{fig:desorption_temperatures}
\end{figure}

Here, we suggest that the experimental desorption temperature can be approximated as the half-life value that matches the typical time scale of the experiment. \figref{fig:desorption_temperatures} shows the distribution of desorption temperatures obtained from \Eqref{eq:desorption} and \Eqref{eq:desorption_halflives} with our distribution of binding energies without ZPE and assuming a half-life time of $\tau=1$~$\mu$s (as approximated time resolution of an experiment). Note that the desorption temperature distribution is rather broad even if a fixed time $\tau$ is assumed. The mean desorption temperature obtained in this way is $28.4$~K, which agrees well with the experimental values found by \cite{Minissale2016} using temperature-programmed desorption (TPD) as  $23-30$~K. It is clear that the desorption prcess may also be influenced by diffusion into deeper binding sites and on the occupancy of these binding sites at different surface coverage. This, however goes beyond the scope of this work and will be addressed in the future. 

\begin{table}
	\centering
	\caption{Rate constants and half-lives of desorption for N atoms on ASW. Values in parenthesis are times in ps.}
	\label{tab:half-lives}
    \begin{tabular}{ccc}
    \hline
    $T_\text{s}$ (K) & $k_\text{d}$ (s$^{-1}$) & $\tau$ (s) \\
    \hline
    10 & 2.5\e{-6} & 2.7\e{5} \\
    20 & 1.7\e{3}  & 4.0\e{-4} \\
    30 & 1.5\e{6}  & 4.6\e{-7}\\
    40 & 4.5\e{7}  & 1.6\e{-8}\\
    50 & 3.4\e{8} & 2.0\e{-9}\\
    90 & 1.3\e{10} & 5.5\e{-11} (55)\\
    130 & 5.1\e{10}& 1.4\e{-11} (14)\\
    \hline
    \end{tabular}
\end{table}

\subsection{Kinetic Energy of the Nitrogen Atom During the Trajectory: Cooling}

\begin{figure*}
    \centering
    \begin{minipage}{0.33\linewidth}
        \centering
        \includegraphics[width=\linewidth]{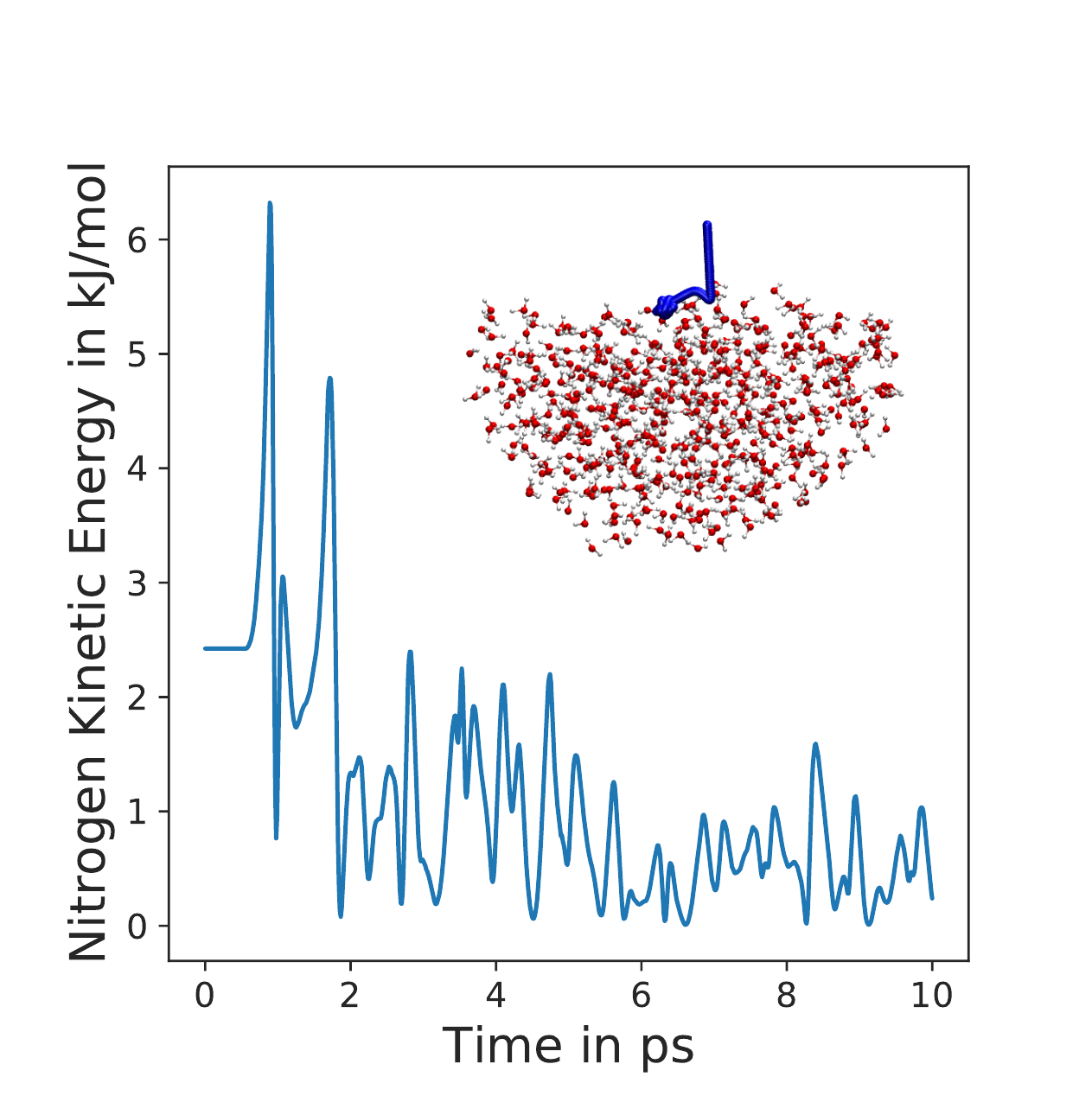}
    \end{minipage}\hfill
    \begin{minipage}{0.33\linewidth}
        \centering
        \includegraphics[width=\linewidth]{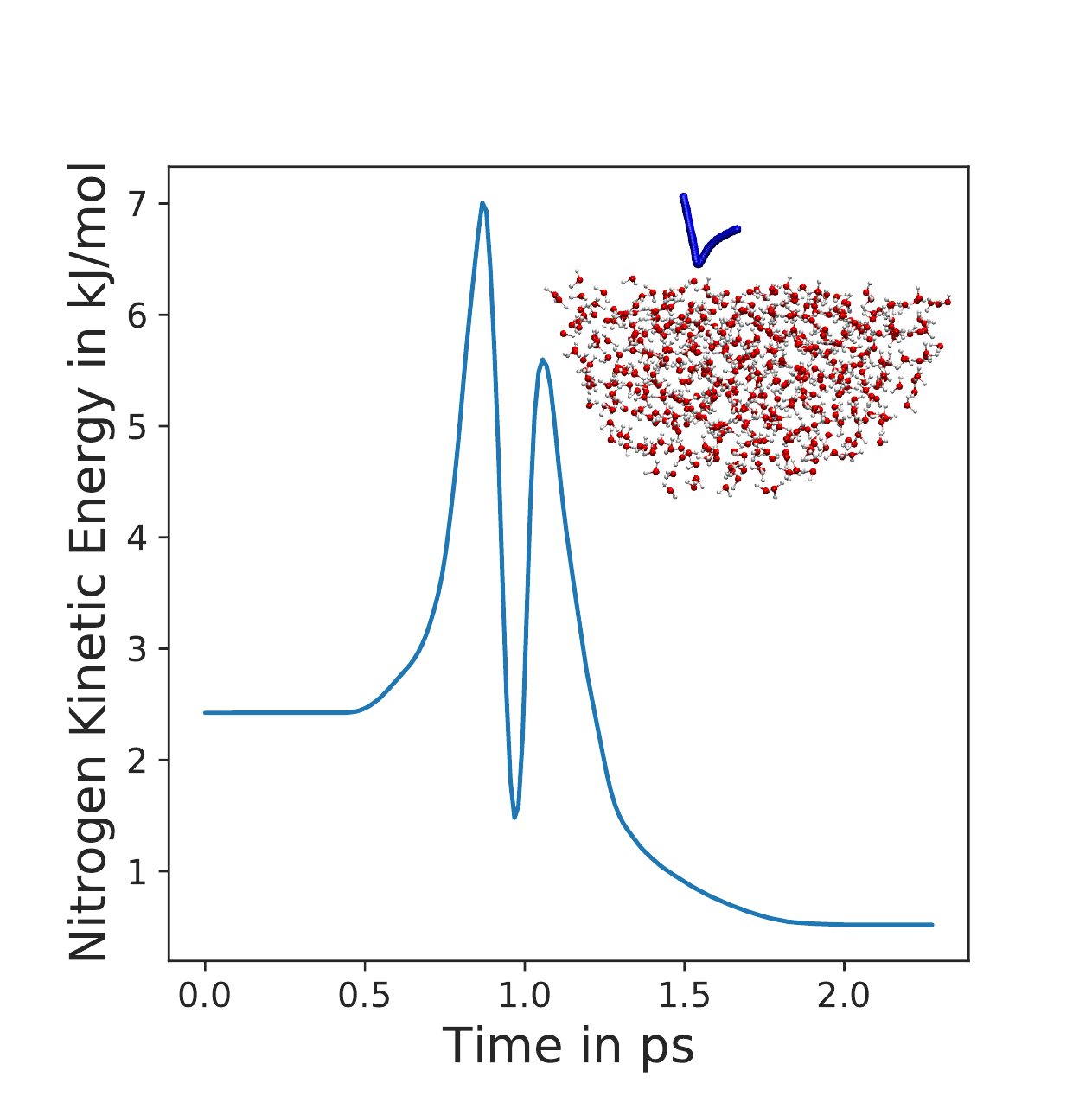}
    \end{minipage}
    \begin{minipage}{0.33\linewidth}
        \centering
        \includegraphics[width=\linewidth]{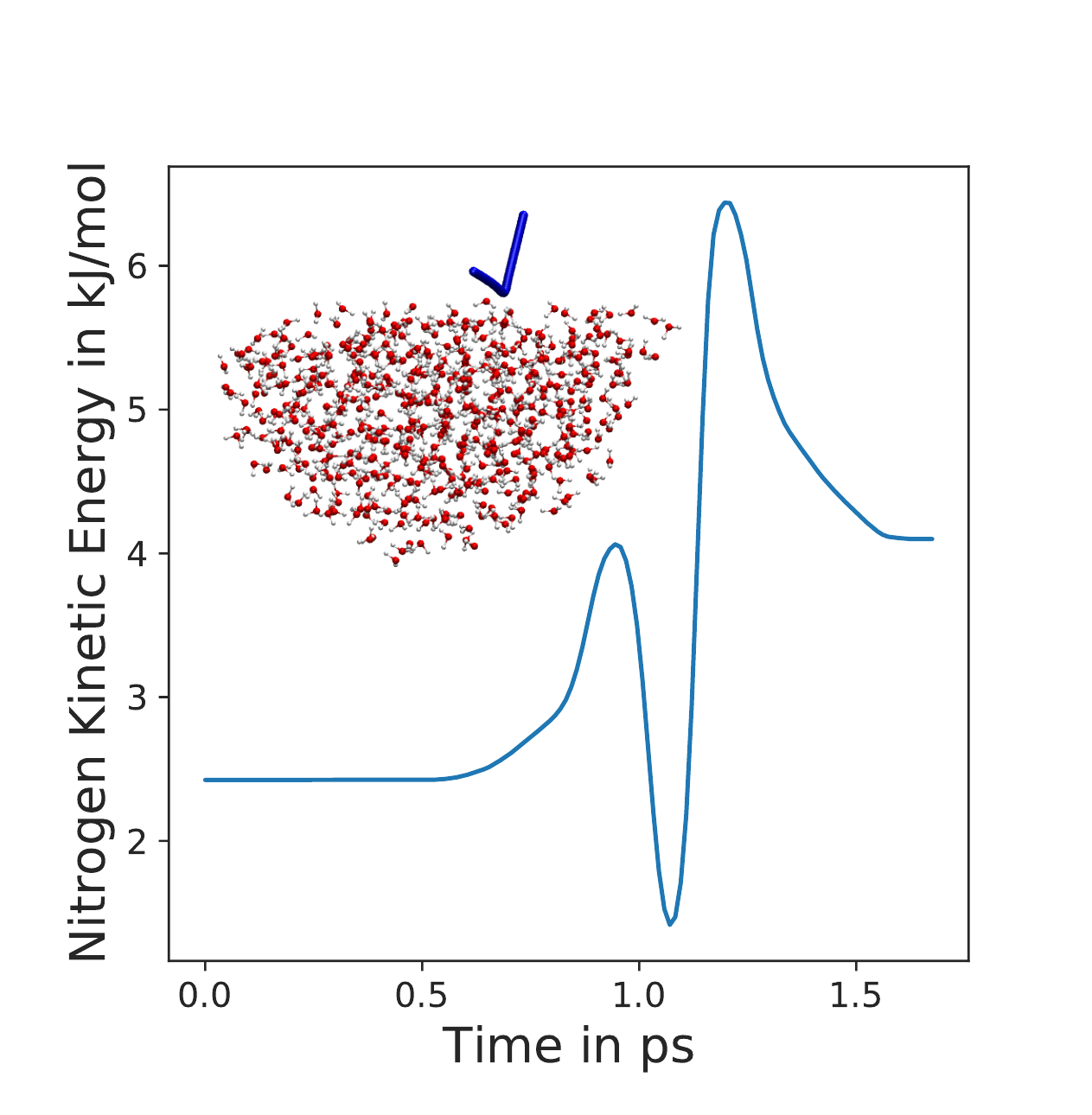}
    \end{minipage}
    \caption{Kinetic energy evolution of three archetypal trajectory types in our simulations, all with $T_\text{s}= 50$~K, $E=2.4$~kJ/mol. \textit{Left:} Sticking ($\theta$=0\degree), \textit{middle:} Bouncing trajectory with net cooling ($\theta$=12\degree), \textit{right:} Bouncing trajectory with net heating ($\theta$=21\degree). }
    \label{fig:cooling_heating}
\end{figure*}

Nitrogen, being a relatively light element with small binding energy with water, is expected to have a sticking coefficient less than unity, even at cryogenic surface temperatures. While this expectation is confirmed by our simulations, see \secref{sec:sticking}, at surface temperatures of $10$~K or $50$~K, the sticking coefficient is only slightly below unity. This can be caused by a rapid cooling of the adsorbed nitrogen atom. To investigate such a cooling process, the trajectories are analyzed in detail. 

\figref{fig:cooling_heating} shows the kinetic energy of the nitrogen atom for selected trajectories at an intermediate surface temperature $T_\text{s}=50$~K and intermediate initial kinetic energy $E=2.1$~kJ/mol. The figure shows bouncing and sticking trajectories. Both types of trajectories start with an acceleration towards the surface as a consequence of the attractive nature of the interaction potential. The height of the first maximum in the kinetic energy corresponds approximately to $E$ plus the binding energy close to the collision point. It occurs at a distance of N to the nearest water molecule, which is close to the equilibrium distance for that configuration. At distances closer than the equilibrium distance, repulsive forces dominate, reducing the kinetic energy.  

After the (first) collision the adsorbate cools down, dissipating the heat into the ice network, see \figref{fig:cooling_heating} (\textit{Left} and \textit{Middle}). Whether a trajectory becomes a sticking or a bouncing event depends on the remaining energy after the collision. If the adsorbate has sufficient energy to leave the potential well it will escape, see \figref{fig:cooling_heating} (\textit{Middle} and \textit{Right}). In such a bouncing situation, the difference between the initial energy and the final energy can be either positive or negative. In the first case, the atom suddenly cools down on the surface during the collision event, see \figref{fig:cooling_heating} (\textit{Middle}). In the second case, the thermal fluctuations of the ice increase the kinetic energy of the N atom, see \figref{fig:cooling_heating} (\textit{Right}). This last situation occurs rarely.

In sticking events, the structure of the kinetic energy plot, \figref{fig:cooling_heating} (\textit{Left}), is significantly different from the bouncing ones. We observe that after the first collision, the first maximum in the kinetic energy, the atom cools down and tries to escape the potential. Since the remaining energy after the collision is not enough, the particle is accelerated again towards the surface, with a lower kinetic energy. This behavior is repeated a few times with increasingly  lower kinetic energy until the adsorbate is thermalized at temperature close to the initial temperature of the ice. 

The time scale of our simulations ($10$~ps) is probably too short for a full thermal equilibration. However, the time scale of this cooling period is essential for the reactivity of the adsorbate. Collisions of the hot adsorbate with surface atoms may allow chemical reactions with significant barriers to happen.

\begin{figure}
    \centering
    \includegraphics[width=\linewidth]{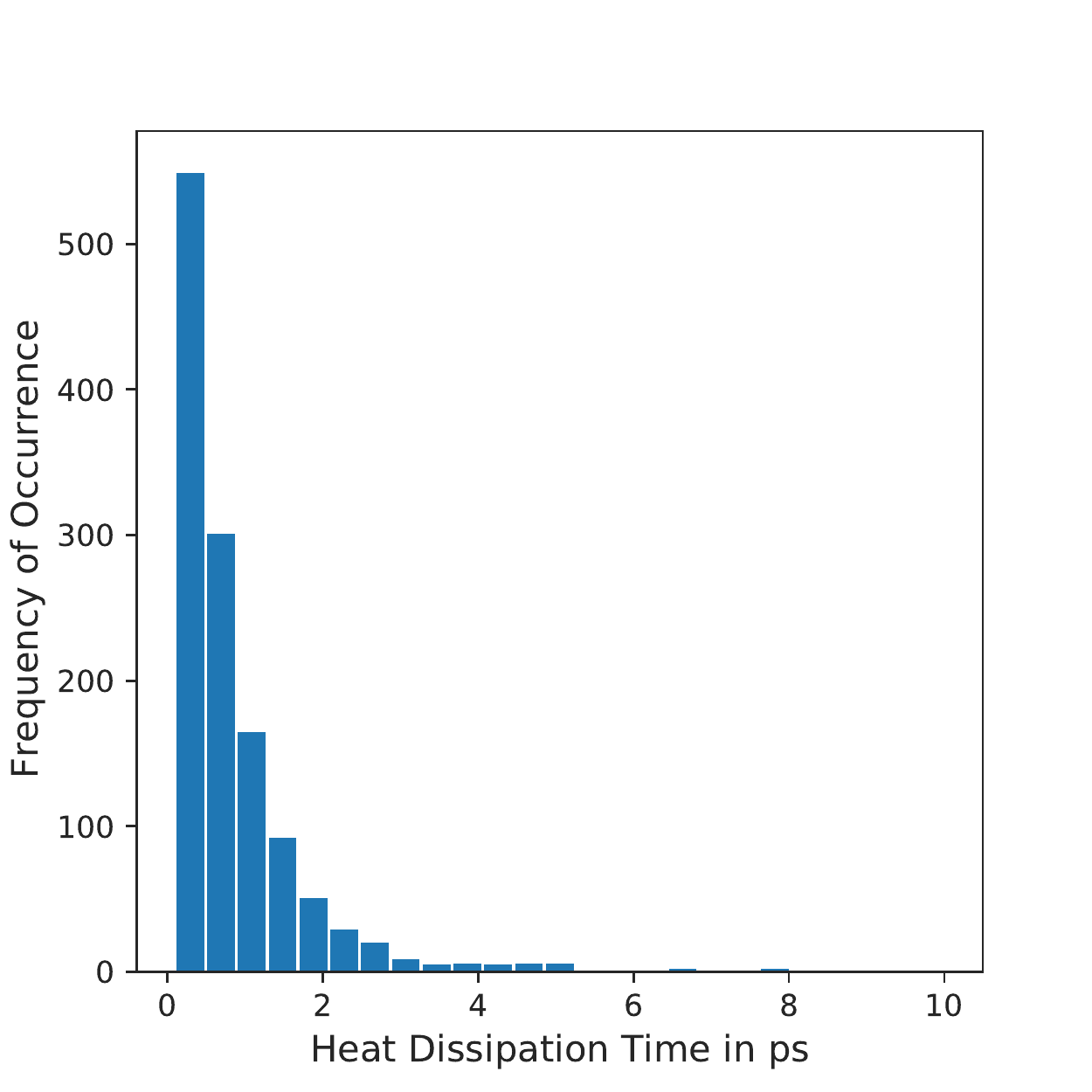}
    \caption{Histogram of the heat dissipation live-times of the nitrogen atom for a ASW surface at $10$~K and the incoming energies depicted in \secref{sec:sticking}. The distribution has a mean of about $0.99$ ps.}
    \label{fig:histogram_dissipation}
\end{figure}

It is possible to estimate the time scale of the energy dissipation by fitting an exponential decay function of the form 
\begin{equation} \label{eq:dissipation}
    E(t)=c_{1}+c_{2}e^{-k_{1}t}
\end{equation}
to the nitrogen kinetic energy in sticking trajectories. 
In \Eqref{eq:dissipation} $c_{1}$ and $c_{2}$ are the fitting constants and $k_{1}$ represents a pseudo rate coefficient for the heat dissipation. The initial time, $t_{0}$, for the fit is defined as the time of collision, the first maximum in the kinetic energy. Energy dissipation is again a uni-molecular process for which decay times (as half-lives) can be estimated according to $\ln{2}/k_1$.
We have computed these decay times for all of our 1546 sticking trajectories at $10$~K. The results are presented as a histogram in \figref{fig:histogram_dissipation}. Some fits fail because of thermal fluctuations that preclude a sufficiently smooth decay of the kinetic energy within 10~ps. These lead to very high or very low decay times, which lie out of the bounds of \figref{fig:histogram_dissipation}. From the figure, we have determined that most of the energy dissipation takes place within 1--2~ps. The cooling procedure is, therefore, very efficient in ASW. Even though we cannot ensure complete thermalization, from the presented results we confirm that it is a likely and fast outcome after a sticking event.

The analysis here was presented for $T_\text{s}=10$~K. For higher surface temperatures, the heat dissipation is more difficult to quantify because of the stronger thermal fluctuations. However, for those trajectories that could be sufficiently fitted by \Eqref{eq:dissipation} we found similar heat dissipation time scales to those for $T_\text{s}=10$~K.

Again, in this work, we have encountered that the probability for sticking (or, more rigorously, spontaneous sticking) is always non-zero, even at high surface temperature, where desorption would eventually occur. However, the time scales differ. Sticking and the subsequent energy dissipation happen within 1--2~ps. Desorption after thermalization occurs at the ns to s time scale, depending on the temperature, see \tabref{tab:half-lives}. We emphasize here that this gap opens up a time window before desorption, for the atom to react. Incorporating this knowledge into astrochemical models will require both, explicit knowledge of the diffusivity of the adsorbate on the substrate and of the rate of thermal desorption. We have provided data on the latter in this work. We are currently working on the modelling of thermal diffusion to complete the input for surface models.

Lastly, it is important to remark that there are several intermediate situations between the energy profiles presented here, attributed to the particular distribution of binding sites on the ice and the initial incoming energy of the adsorbate. For example, in the rare deep minima, the cooling can be much faster with a lower degree of thermal oscillation. Other situations can be found when definitive bouncing happens after multiple individual collisions with the surface. 

\section{Astrophysical Implications and Conclusions}

Nitrogen interstellar chemistry is deeply affected by physical processes before chemical reactions. As it was studied by \cite{Shimonishi2018}, binding energies significantly affect the chemistry of nitrogen in molecular clouds. Combining \textit{ab-initio} and rate equation approaches, they recalculated the N-ASW adsorption energies and N-bearing molecular abundances. Their calculations predicted a predominance of \ce{N2} in the modeled astronomical object, even inhibiting the formation of \ce{NH3}, up to $\sim$ 4.5\e{4} years after its formation. This dichotomy has been debated in the literature \citep{Snow2004, Daranlot2012}. However, exhaustive modelling of these processes requires additional data on several quantities. With this work, we have provided data on sticking and desorption. Our values for the binding energies are in very close agreement with the ones obtained in \cite{Shimonishi2018} and in good agreement with the experimental results of \cite{Minissale2016} (by a factor of 1.8). The dispersion in our results is significant, indicating that the low binding energy of nitrogen is strongly dependent on the adsorption sites. Our main astrophysical conclusions agree with the alternative scenario, including \ce{N2} lock, or at least do not provide any conclusive evidence against it. The relatively large value of the sticking coefficient $S_{T}$ at low temperatures, characteristic of dense molecular clouds, guarantees that provided a small kinetic temperature of the gas, all nitrogen atoms will adhere to the surface. Diffusivity of nitrogen on water ice is a significant quantity to unravel the conundrum.

Sticking coefficients at surface temperatures higher than $10$~K are always lower than unity, due to the small value of the binding energy. This has implications on the chemistry of warmer objects, such as translucent clouds, where ice mantles start to be formed \citep{Cazaux2011}.  In order to model nitrogen chemistry in these objects, especially surface chemistry on grains, it is advisable to correct the models by the quantities provided in this work. The same logic can be applied to the chemistry of other warmer regions, such as protoplanetary disks with an average temperature in the outer parts of the disk in between 50 and 150~K \citep{Boss1998} that allow rich chemistry to be initiated \citep{Henning2013}. \cite{Schwarz2014} provide an interesting discussion about how the initial abundances of N on nitrogen-bearing molecules affect the spatial distribution of N in the disk. We suggest that our values of sticking coefficients and binding energies can improve astrochemical models, especially considering that sticking and diffusion are possible above the desorption temperature in a time-frame of tens of ps to $\mu$s and longer. In cold conditions, such as the ones in molecular clouds, it is safe to consider a sticking coefficient of the unity for most cases. For systems at higher surface and gas temperatures, we recommend to employ values of the sticking coefficients from \Tabref{tab:sticking_coefficients}. Concerning binding energies, it is advisable to employ the mean value from the distribution, including ZPE (-345.6~K), and if possible the whole distribution of binding energies.

Another quantity determined during this study is the experimentally measurable desorption temperature for N on water ice. We emphasize here that there is no unique definition of desorption temperature and that chemical surface reactions are possible at temperatures higher than the desorption temperature, due to the higher available thermal energy for diffusion. The definition of desorption temperature depends on the temporal resolution of the measuring technique. The exact importance of diffusion above an experimental temperature varies according to the mobility of the adsorbate before desorption. 

Energy dissipation and non-thermal diffusion are currently important aspects of interstellar surface chemistry that have been studied in recent years with implications to the chemistry of dense clouds \citep{Lamberts2014, Fredon2018} and a focus on exothermicity of chemical reactions. Here we presented the energy dissipation after a collision leading to physisorption. Nitrogen atoms dissipate their initial kinetic energy plus their binding energy within the first picosecond after the collision. The magnitude of the local heating may be lower than the one coming from an exothermic chemical reaction but still, such short time scales essentially preclude 
hot-atom reactivity caused by a collision event with the ASW surface. 

From a technical point of view, we would like to draw attention to the interatomic potential that we have constructed. The neural network potential used in this work has the accuracy of underlying DFT methods at a cost close to empirical force fields. A measure of the goodness of the potential comes from the estimation of the binding energies, which are in close agreement with experiments and explicit DFT calculations alike. We have also determined the degree of deviation of our model in Appendix \ref{sec:extrapolation}. This methodology needs an intense iterative construction of the training set, which can be time-consuming. We are currently applying the GM-NN approach to the simulation of other complex surface processes, which eventually would lead to an overall better description of surface processes in the ISM.

\section*{Acknowledgements}

 We thank the Deutsche Forschungsgemeinschaft (DFG, German Research Foundation) for supporting this work by funding EXC 2075 - 390740016 under Germany's Excellence Strategy. We acknowledge the support by the Stuttgart Center for Simulation Science (SimTech) and the European Union's Horizon 2020 research and innovation programme (grant agreement No. 646717, TUNNELCHEM). We also like to acknowledge the support by the Institute for Parallel and Distributed Systems (IPVS) of the University of Stuttgart and by the state of Baden-W\"urttemberg through the bwHPC consortium for providing computer time. G. M. thanks the Alexander von Humboldt Foundation for their patronage with a postdoctoral fellowship. V. Z. acknowledges financial support received in the form of a PhD scholarship from the Studienstiftung  des  Deutschen  Volkes (German National Academic Foundation).

\section*{Data availability}

The data obtained in this article will be shared on reasonable request to the corresponding author.




\bibliographystyle{mnras}
\bibliography{example} 

\begin{thebibliography}{}
\makeatletter
\relax
\def\mn@urlcharsother{\let\do\@makeother \do\$\do\&\do\#\do\^\do\_\do\%\do\~}
\def\mn@doi{\begingroup\mn@urlcharsother \@ifnextchar [ {\mn@doi@}
  {\mn@doi@[]}}
\def\mn@doi@[#1]#2{\def\@tempa{#1}\ifx\@tempa\@empty \href
  {http://dx.doi.org/#2} {doi:#2}\else \href {http://dx.doi.org/#2} {#1}\fi
  \endgroup}
\def\mn@eprint#1#2{\mn@eprint@#1:#2::\@nil}
\def\mn@eprint@arXiv#1{\href {http://arxiv.org/abs/#1} {{\tt arXiv:#1}}}
\def\mn@eprint@dblp#1{\href {http://dblp.uni-trier.de/rec/bibtex/#1.xml}
  {dblp:#1}}
\def\mn@eprint@#1:#2:#3:#4\@nil{\def\@tempa {#1}\def\@tempb {#2}\def\@tempc
  {#3}\ifx \@tempc \@empty \let \@tempc \@tempb \let \@tempb \@tempa \fi \ifx
  \@tempb \@empty \def\@tempb {arXiv}\fi \@ifundefined
  {mn@eprint@\@tempb}{\@tempb:\@tempc}{\expandafter \expandafter \csname
  mn@eprint@\@tempb\endcsname \expandafter{\@tempc}}}

\bibitem[\protect\citeauthoryear{Abadi et~al.,}{Abadi et~al.}{2015}]{TF15}
Abadi M.,  et~al., 2015, {TensorFlow}: Large-Scale Machine Learning on
  Heterogeneous Systems, \url {https://www.tensorflow.org/}

\bibitem[\protect\citeauthoryear{Balasubramani et~al.,}{Balasubramani
  et~al.}{2020}]{Turbomole}
Balasubramani S.~G.,  et~al., 2020, \mn@doi [J. Chem. Phys.]
  {10.1063/5.0004635}, 152, 184107

\bibitem[\protect\citeauthoryear{Bannwarth, Ehlert  \& Grimme}{Bannwarth
  et~al.}{2019}]{Bannwarth2019}
Bannwarth C.,  Ehlert S.,   Grimme S.,  2019, \mn@doi [J. Chem. Theory Comput.]
  {10.1021/acs.jctc.8b01176}, 15, 1652

\bibitem[\protect\citeauthoryear{Behler \& Parrinello}{Behler \&
  Parrinello}{2007}]{Behler07}
Behler J.,  Parrinello M.,  2007, \mn@doi [Phys. Rev. Lett.]
  {10.1103/PhysRevLett.98.146401}, 98, 146401

\bibitem[\protect\citeauthoryear{Boss}{Boss}{1998}]{Boss1998}
Boss A.~P.,  1998, \mn@doi [Annu. Rev. Earth Planet. Ciencia]
  {10.1146/annurev.earth.26.1.53}, 26, 53

\bibitem[\protect\citeauthoryear{Buch \& Zhang}{Buch \& Zhang}{1991}]{Buch1991}
Buch V.,  Zhang Q.,  1991, \mn@doi [ApJ] {10.1086/170537}, 379, 647

\bibitem[\protect\citeauthoryear{Cazaux, Caselli  \& Spaans}{Cazaux
  et~al.}{2011}]{Cazaux2011}
Cazaux S.,  Caselli P.,   Spaans M.,  2011, \mn@doi [ApJ]
  {10.1088/2041-8205/741/2/L34}, 741, L34

\bibitem[\protect\citeauthoryear{Cuppen, Walsh, Lamberts, Semenov, Garrod,
  Penteado  \& Ioppolo}{Cuppen et~al.}{2017}]{Cuppen2017}
Cuppen H.~M.,  Walsh C.,  Lamberts T.,  Semenov D.,  Garrod R.~T.,  Penteado
  E.~M.,   Ioppolo S.,  2017, \mn@doi [Space Sci. Rev]
  {10.1007/s11214-016-0319-3}, 212

\bibitem[\protect\citeauthoryear{Daranlot, Hincelin, Bergeat, Costes, Loison,
  Wakelam  \& Hickson}{Daranlot et~al.}{2012}]{Daranlot2012}
Daranlot J.,  Hincelin U.,  Bergeat A.,  Costes M.,  Loison J.~C.,  Wakelam V.,
    Hickson K.~M.,  2012, \mn@doi [Proc. Nat. Acad. Sci. U.S.A]
  {10.1073/pnas.1200017109}, 109, 10233

\bibitem[\protect\citeauthoryear{Enrique-Romero, Rimola, Ceccarelli  \&
  Balucani}{Enrique-Romero et~al.}{2016}]{Enrique-Romero2016}
Enrique-Romero J.,  Rimola A.,  Ceccarelli C.,   Balucani N.,  2016, \mn@doi
  [MNRAS] {10.1093/mnrasl/slw031}, 459, L6

\bibitem[\protect\citeauthoryear{Fedoseev, Chuang, Ioppolo, Qasim, van Dishoeck
   \& Linnartz}{Fedoseev et~al.}{2017}]{Fedoseev2017}
Fedoseev G.,  Chuang K.-J.,  Ioppolo S.,  Qasim D.,  van Dishoeck E.~F.,
  Linnartz H.,  2017, \mn@doi [ApJ] {10.3847/1538-4357/aa74dc}, 842, 52

\bibitem[\protect\citeauthoryear{Fredon \& Cuppen}{Fredon \&
  Cuppen}{2018}]{Fredon2018}
Fredon A.,  Cuppen H.~M.,  2018, \mn@doi [Phys. Chem. Chem. Phys.]
  {10.1039/c7cp06136f}, 20, 5569

\bibitem[\protect\citeauthoryear{Frisch et~al.,}{Frisch et~al.}{2016}]{g16}
Frisch M.~J.,  et~al., 2016, Gaussian˜16 {R}evision {C}.01

\bibitem[\protect\citeauthoryear{Grimme, Brandenburg, Bannwarth  \&
  Hansen}{Grimme et~al.}{2015}]{Grimme2015}
Grimme S.,  Brandenburg J.~G.,  Bannwarth C.,   Hansen A.,  2015, \mn@doi [J.
  Chem. Phys.] {10.1063/1.4927476}, 143, 054107

\bibitem[\protect\citeauthoryear{Grimme, Bannwarth  \& Shushkov}{Grimme
  et~al.}{2017}]{Grimme2017}
Grimme S.,  Bannwarth C.,   Shushkov P.,  2017, \mn@doi [J. Chem. Theory
  Comput.] {10.1021/acs.jctc.7b00118}, 13, 1989

\bibitem[\protect\citeauthoryear{Henning \& Semenov}{Henning \&
  Semenov}{2013}]{Henning2013}
Henning T.,  Semenov D.,  2013, \mn@doi [Chem. Rev.] {10.1021/cr400128p}, 113,
  9016

\bibitem[\protect\citeauthoryear{{Hjorth Larsen} et~al.,}{{Hjorth Larsen}
  et~al.}{2017}]{HjorthLarsen2017}
{Hjorth Larsen} A.,  et~al., 2017, \mn@doi [J. Condens. Matter Phys.]
  {10.1088/1361-648X/aa680e}, 29, 273002

\bibitem[\protect\citeauthoryear{Humphrey, Dalke  \& Schulten}{Humphrey
  et~al.}{1996}]{VMD}
Humphrey W.,  Dalke A.,   Schulten K.,  1996, \mn@doi [J. Mol. Graph. Model.]
  {10.1016/0263-7855(96)00018-5}, 14, 33

\bibitem[\protect\citeauthoryear{Jenniskens \& Blake}{Jenniskens \&
  Blake}{1994}]{Jenniskens1994}
Jenniskens P.,  Blake D.~F.,  1994, \mn@doi [Science]
  {10.1126/science.11539186}, 265, 753

\bibitem[\protect\citeauthoryear{Jorgensen}{Jorgensen}{1981}]{Jorgensen1981}
Jorgensen W.~L.,  1981, \mn@doi [J. Am. Chem. Soc.] {10.1021/ja00392a016}, 103,
  335

\bibitem[\protect\citeauthoryear{Kobayashi, Hidaka, Lamberts, Hama, Kawakita,
  K{\"{a}}stner  \& Watanabe}{Kobayashi et~al.}{2017}]{Kobayashi2017}
Kobayashi H.,  Hidaka H.,  Lamberts T.,  Hama T.,  Kawakita H.,  K{\"{a}}stner
  J.,   Watanabe N.,  2017, \mn@doi [ApJ] {10.3847/1538-4357/837/2/155}, 837,
  155

\bibitem[\protect\citeauthoryear{Lamberts, de Vries  \& Cuppen}{Lamberts
  et~al.}{2014}]{Lamberts2014}
Lamberts T.,  de Vries X.,   Cuppen H.~M.,  2014, \mn@doi [Faraday Discuss.]
  {10.1039/C3FD00136A}, 168, 327

\bibitem[\protect\citeauthoryear{Lamberts, Markmeyer, Kolb  \&
  K{\"{a}}stner}{Lamberts et~al.}{2019}]{Lamberts2019}
Lamberts T.,  Markmeyer M.~N.,  Kolb F.~J.,   K{\"{a}}stner J.,  2019, \mn@doi
  [ACS. Earth. Space. Chem.] {10.1021/acsearthspacechem.9b00029}, 3, 958

\bibitem[\protect\citeauthoryear{Linnartz, Ioppolo  \& Fedoseev}{Linnartz
  et~al.}{2015}]{Linnartz2015}
Linnartz H.,  Ioppolo S.,   Fedoseev G.,  2015, \mn@doi [Int. Rev. Phys. Chem.]
  {10.1080/0144235X.2015.1046679}

\bibitem[\protect\citeauthoryear{Masuda, Takahashi  \& Mukai}{Masuda
  et~al.}{1998}]{Masuda1998}
Masuda K.,  Takahashi J.,   Mukai T.,  1998, A\&A, 330, 773

\bibitem[\protect\citeauthoryear{Meisner, Lamberts  \& K{\"{a}}stner}{Meisner
  et~al.}{2017}]{Meisner2017}
Meisner J.,  Lamberts T.,   K{\"{a}}stner J.,  2017, \mn@doi [ACS. Earth.
  Space. Chem.] {10.1021/acsearthspacechem.7b00052}, 1, 399

\bibitem[\protect\citeauthoryear{Metz, K{\"{a}}stner, Sokol, Keal  \&
  Sherwood}{Metz et~al.}{2014}]{Chemshell}
Metz S.,  K{\"{a}}stner J.,  Sokol A.~A.,  Keal T.~W.,   Sherwood P.,  2014,
  \mn@doi [Wiley Interdiscip. Rev. Comput. Mol. Sci.] {10.1002/wcms.1163}, 4,
  101

\bibitem[\protect\citeauthoryear{{Minissale, M.}, {Congiu, E.}  \& {Dulieu,
  F.}}{{Minissale, M.} et~al.}{2016}]{Minissale2016}
{Minissale, M.} {Congiu, E.}  {Dulieu, F.} 2016, \mn@doi [A\&A]
  {10.1051/0004-6361/201526702}, 585, A146

\bibitem[\protect\citeauthoryear{Molpeceres \& K{\"{a}}stner}{Molpeceres \&
  K{\"{a}}stner}{2020}]{Molpeceres2020}
Molpeceres G.,  K{\"{a}}stner J.,  2020, \mn@doi [Phys. Chem. Chem. Phys.]
  {10.1039/d0cp00250j}, 22, 7552

\bibitem[\protect\citeauthoryear{Molpeceres, Rimola, Ceccarelli, K{\"{a}}stner,
  Ugliengo  \& Mat{\'{e}}}{Molpeceres et~al.}{2019}]{Molpeceres2019}
Molpeceres G.,  Rimola A.,  Ceccarelli C.,  K{\"{a}}stner J.,  Ugliengo P.,
  Mat{\'{e}} B.,  2019, \mn@doi [MNRAS] {10.1093/mnras/sty3024}, 482, 5389

\bibitem[\protect\citeauthoryear{Oba, Takano, Naraoka, Watanabe  \& Kouchi}{Oba
  et~al.}{2019}]{Oba2019}
Oba Y.,  Takano Y.,  Naraoka H.,  Watanabe N.,   Kouchi A.,  2019, \mn@doi
  [Nature Communications] {10.1038/s41467-019-12404-1}, 10

\bibitem[\protect\citeauthoryear{{\"{O}}berg}{{\"{O}}berg}{2016}]{Oberg2016}
{\"{O}}berg K.~I.,  2016, \mn@doi [Chem. Rev.] {10.1021/acs.chemrev.5b00694},
  116, 9631

\bibitem[\protect\citeauthoryear{Phillips et~al.,}{Phillips
  et~al.}{2005}]{Phillips2005}
Phillips J.~C.,  et~al., 2005, \mn@doi [J. Comput. Chem.] {10.1002/jcc.20289},
  26, 1781

\bibitem[\protect\citeauthoryear{Potapov, Theul{\'{e}}, J{\"{a}}ger  \&
  Henning}{Potapov et~al.}{2019}]{Potapov2019}
Potapov A.,  Theul{\'{e}} P.,  J{\"{a}}ger C.,   Henning T.,  2019, \mn@doi
  [ApJ] {10.3847/2041-8213/ab2538}, 878, L20

\bibitem[\protect\citeauthoryear{Potapov, J{\"{a}}ger  \& Henning}{Potapov
  et~al.}{2020}]{Potapov2020}
Potapov A.,  J{\"{a}}ger C.,   Henning T.,  2020, \mn@doi [ApJ]
  {10.3847/1538-4357/ab86b5}, 894, 110

\bibitem[\protect\citeauthoryear{Qasim, Fedoseev, Lamberts, Chuang, He,
  Ioppolo, K{\"{a}}stner  \& Linnartz}{Qasim et~al.}{2019}]{Qasim2019}
Qasim D.,  Fedoseev G.,  Lamberts T.,  Chuang K.~J.,  He J.,  Ioppolo S.,
  K{\"{a}}stner J.,   Linnartz H.,  2019, \mn@doi [ACS Earth. Space. Chem.]
  {10.1021/acsearthspacechem.9b00062}, 3, 986

\bibitem[\protect\citeauthoryear{Reddi, Kale  \& Kumar}{Reddi
  et~al.}{2019}]{Reddi18}
Reddi S.~J.,  Kale S.,   Kumar S.,  2019, arXiv, 1904.09237 [cs.LG]

\bibitem[\protect\citeauthoryear{Rimola, Skouteris, Balucani, Ceccarelli,
  Enrique-Romero, Taquet  \& Ugliengo}{Rimola et~al.}{2018}]{Rimola2018}
Rimola A.,  Skouteris D.,  Balucani N.,  Ceccarelli C.,  Enrique-Romero J.,
  Taquet V.,   Ugliengo P.,  2018, \mn@doi [ACS Earth. Space. Chem.]
  {10.1021/acsearthspacechem.7b00156}, 2, 720

\bibitem[\protect\citeauthoryear{Schwarz \& Bergin}{Schwarz \&
  Bergin}{2014}]{Schwarz2014}
Schwarz K.~R.,  Bergin E.~A.,  2014, \mn@doi [ApJ]
  {10.1088/0004-637X/797/2/113}, 797, 113

\bibitem[\protect\citeauthoryear{Seabold \& Perktold}{Seabold \&
  Perktold}{2010}]{Statmodels}
Seabold S.,  Perktold J.,  2010, in 9th Python in Science Conference.

\bibitem[\protect\citeauthoryear{Settles}{Settles}{2009}]{settles.tr09}
Settles B.,  2009, Computer Sciences Technical Report~1648, Active Learning
  Literature Survey.
University of Wisconsin--Madison

\bibitem[\protect\citeauthoryear{Sherwood et~al.,}{Sherwood
  et~al.}{2003}]{Sherwood2003}
Sherwood P.,  et~al., 2003, \mn@doi [J. Mol. Struc-THEOCHEM]
  {10.1016/s0166-1280(03)00285-9}, 632, 1

\bibitem[\protect\citeauthoryear{Shimonishi, Nakatani, Furuya  \&
  Hama}{Shimonishi et~al.}{2018}]{Shimonishi2018}
Shimonishi T.,  Nakatani N.,  Furuya K.,   Hama T.,  2018, \mn@doi [ApJ]
  {10.3847/1538-4357/aaaa6a}, 855, 27

\bibitem[\protect\citeauthoryear{Snow}{Snow}{2004}]{Snow2004}
Snow T.~P.,  2004, \mn@doi [Nature] {10.1038/429615a}, 429, 615

\bibitem[\protect\citeauthoryear{Snow \& McCall}{Snow \&
  McCall}{2006}]{Snow2006}
Snow T.~P.,  McCall B.~J.,  2006, \mn@doi [Annu. Rev. Astron. Astrophys.]
  {10.1146/annurev.astro.43.072103.150624}, 44, 367

\bibitem[\protect\citeauthoryear{Stewart}{Stewart}{2013}]{Stewart2013}
Stewart J. J.~P.,  2013, J. Mol. Model., 19, 1

\bibitem[\protect\citeauthoryear{Todorov, Smith, Trachenko  \& Dove}{Todorov
  et~al.}{2006}]{dlpoly}
Todorov I.~T.,  Smith W.,  Trachenko K.,   Dove M.~T.,  2006, \mn@doi [J.
  Mater. Chem.] {10.1039/B517931A}, 16, 1911

\bibitem[\protect\citeauthoryear{Wakelam, Loison, Mereau  \& Ruaud}{Wakelam
  et~al.}{2017}]{Wakelam2017}
Wakelam V.,  Loison J.~C.,  Mereau R.,   Ruaud M.,  2017, \mn@doi [Mol.
  Astrophys.] {10.1016/j.molap.2017.01.002}, 6, 22

\bibitem[\protect\citeauthoryear{Watanabe \& Kouchi}{Watanabe \&
  Kouchi}{2008}]{Watanabe2008}
Watanabe N.,  Kouchi A.,  2008, \mn@doi [Prog. Surf. Sci.]
  {10.1016/j.progsurf.2008.10.001}, 83, 439

\bibitem[\protect\citeauthoryear{Werner, Knowles, Knizia, Manby  \&
  Sch{\"{u}}tz}{Werner et~al.}{2012}]{Molpro}
Werner H.-J.,  Knowles P.~J.,  Knizia G.,  Manby F.~R.,   Sch{\"{u}}tz M.,
  2012, WIREs Comput Mol Sci, 2, 242

\bibitem[\protect\citeauthoryear{Werner et~al.,}{Werner
  et~al.}{2015}]{molpro2015}
Werner H.-J.,  et~al., 2015, MOLPRO, version 2015.1, a package of ab initio
  programs

\bibitem[\protect\citeauthoryear{Zaverkin \& K{\"{a}}stner}{Zaverkin \&
  K{\"{a}}stner}{2020}]{Zaverkin2020}
Zaverkin V.,  K{\"{a}}stner J.,  2020, \mn@doi [J. Chem. Theory Comput.]
  {10.1039/d0cp00250j}, 16, 5410

\makeatother
\end{thebibliography}



\appendix

\section{Density Functional Exchange \& Correlation Benchmark} \label{sec:benchmark}

Benchmark of the refinement method, \textbf{(2)} that we employed for the training data generation and, thus, the underlying method for our dynamics simulations, has been done in a similar context as in our previous work \citep{Molpeceres2020}. At different constant radii from the center of mass of a single water molecule, we place nitrogen atoms uniformly distributed on a spherical surface defined by a Fibonacci lattice. Each of these arrangements defines a configuration from which we extract single point energies of a reference method, CCSD(T)-F12/cc-pVTZ-F12 (calculated using Molpro2015 \cite{Molpro, molpro2015}) and also single-points of the methods under examination. The functionals we considered are B3LYP, BHLYP, BP86, PBE, PBE0, TPSSH (all of them accompanied by a def2-TZVP basis set, and D3BJ dispersion corrected) and PBEh-3c (def2-mSVP). All DFT calculations have been carried out with the Turbomole package v.7.4.1 \citep{Turbomole}. We have sampled the energy of 25 configurations per radius for a total of 4 radii covering a range of distances ($2.0$, $3.0$, $4.0$ \& $5.0$~\AA). Then, the interaction energies are calculated as $E = E_{\ce{H2O}/N}-(E_{\ce{H2O}}+E_{N})$. The results for the benchmark are presented in \Tabref{tab:benchmark}, in which the mean absolute error (MAE) of the interaction energies of the different combinations of functionals/configurations with respect the CCSD(T)-F12 values is presented.

We have observed that the PBEh-3c/def2-mSVP is the most consistent method at long and short ranges, likely due to the systematic geometric counterpoise correction. Some functionals excel at short range, some at long range, but the most consistent is the one we have chosen. Additionally, it is computationally cheaper when compared with conventional hybrids with a triple $\zeta$ basis. As a final note, interaction energies for a radius of the sphere of $2$~\AA~are higher because they correspond to short-range repulsive interactions that are seldom or never sampled in a molecular dynamics run.

\begin{table}
	\centering
	\caption{Benchmark of DFT functionals. Numbers represent the mean absolute deviation from the CCSD(T)-F12 values in kJ/mol. }
	\label{tab:benchmark}
    \begin{tabular}{ccccc}
    \hline
    & \multicolumn{4}{c}{Radius of the sphere (\AA)} \\
    \hline
    Functional & 2.0 & 3.0 & 4.0 & 5.0 \\
    \hline
    B3LYP & 16.6 & 1.9 & 0.4 & 0.03\\
    BHLYP & 3.4 & 0.6 & 1.7 & 1.85\\
    BP86 &  29.1 & 1.8 & 0.2 & 0.02\\
    PBE &   32.1 & 3.9 & 1.0& 0.13\\
    PBE0 &  16.6 & 1.7 & 0.6 & 0.09  \\
    TPSSH & 17.9 & 1.3 & 0.7 & 0.12  \\
    \textbf{PBEh-3c} & 10.3 &1.7 & 0.2 & 0.03 \\
    \hline
    \end{tabular}
\end{table}

An additional estimation that we can extract from this benchmark is the maximum average interaction energy as a function of the radius of the sphere. It is necessary for defining the cutoff radius used during training of the neural network and subsequent inferences. For $5$~\AA~this interaction energy is about $-21$~K (CCSD(T)-F12 values), i.e., is about 5~\% of the average binding energy (non-ZPE corrected). For the employed cutoff radius ($5.5$~\AA), this average interaction energy is  $-11$~K. It is about $2.5$~\% of the binding energy. We consider a cutoff radius of $5.5$~\AA~to be a good compromise since it allows us to generalize the behavior of the \ce{H2O} ice satisfactorily and at the same time covers the vast majority of the interaction potential of the N-\ce{H2O} pair at long distances.

\begin{figure}
        \centering
        \includegraphics[width=\linewidth]{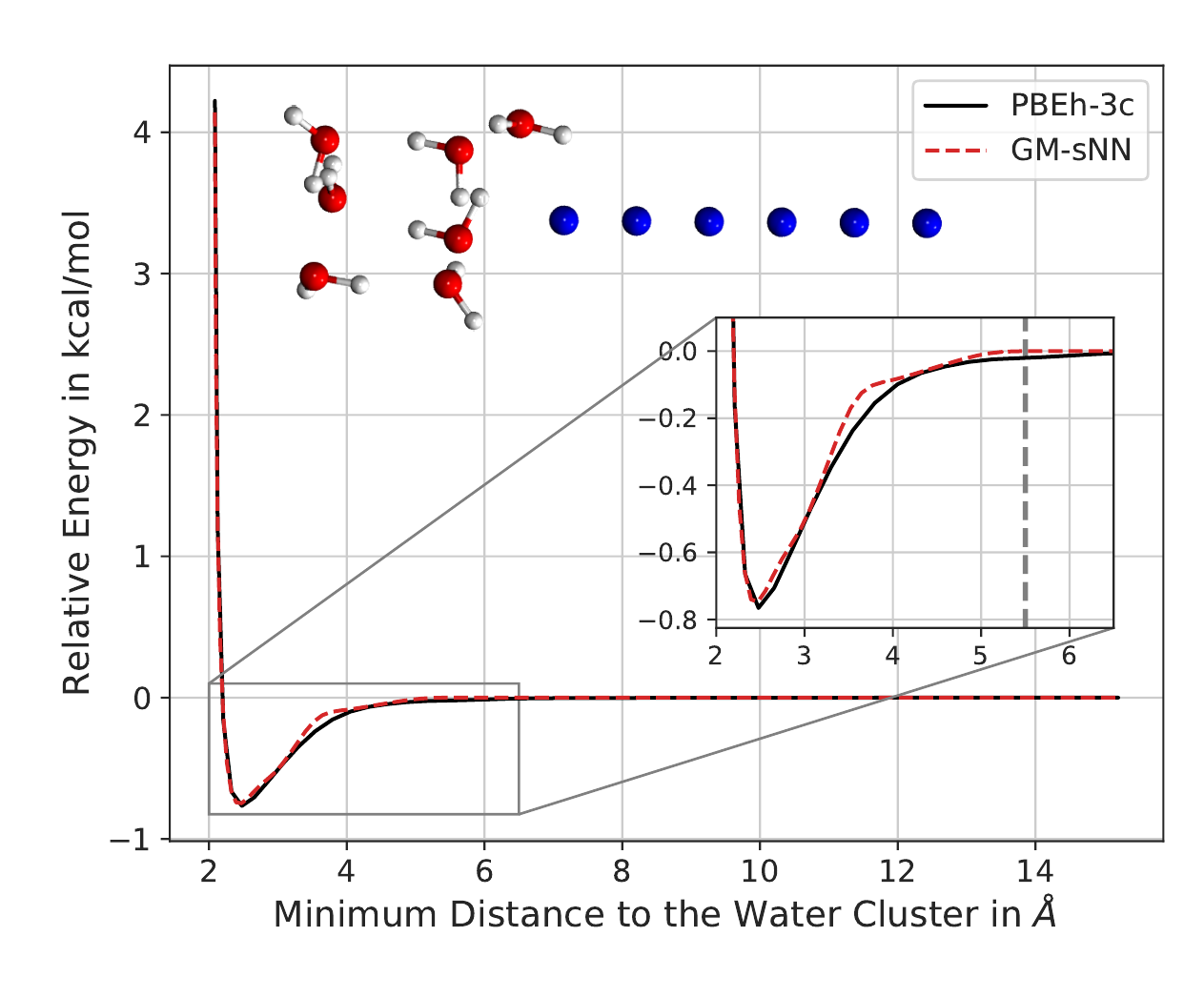}
    \caption{Potential energy profile for the distance of the nitrogen atom to the water cluster, representing the detachment of the nitrogen from the surface. Grey dashed line in the inset represents the cutoff radius of 5.5~{\AA} used during training the GM-sNN model.}
    \label{fig:dft_vs_gmnn}
\end{figure}

Lastly, we have tested the quality of our machine-learned potential (MLP) by performing a rigid potential energy scan. It represents the desorption of the N atom from a seven water cluster. Single point calculations were carried out at equispaced points between the N atom and the center of mass of the cluster (ranging from 3.3 to 18~{\AA}). We used both our MLP and the underlying DFT method (PBEh-3c) for this purpose. The results for this simulation are presented in \figref{fig:dft_vs_gmnn}. Note that the $x$-axis represents the minimum distance from the nitrogen atom to any atom of the cluster. Variations in the GM-sNN energy occur because our training set is not ideal for such small clusters but tailored at the full system.
The GM-sNN potential drops smoothly to zero at the cutoff distance. While this does not reproduce the long-range interactions, the qualitatively correct potential at short distances and the smooth behavior in the cutoff region ensure that the kinetic energy at distances below about 5~{\AA} is well reproduced. This is important for correct dynamics in the bouncing and sticking processes.

\section{Query-By-Committee Approach for Tracking the MLP Performance} \label{sec:extrapolation}

For tracking the performance of the GM-NN model during the collision dynamics the Query-By-Committee (QBC) approach was used \citep{settles.tr09}. The QBC approach employs a committee of models which are trained on the same labeled training data. Each model predicts then the labeling of some query candidate and the instance about which they most disagree is considered the most informative query. Note that usually the QBC approach is used to select data which has to be included into the training set. However, in this work a slightly different application of this approach is used. We use it to identify points along a trajectory that lie outside of the training set.

\begin{figure*}
    \centering
    \includegraphics[width=\linewidth]{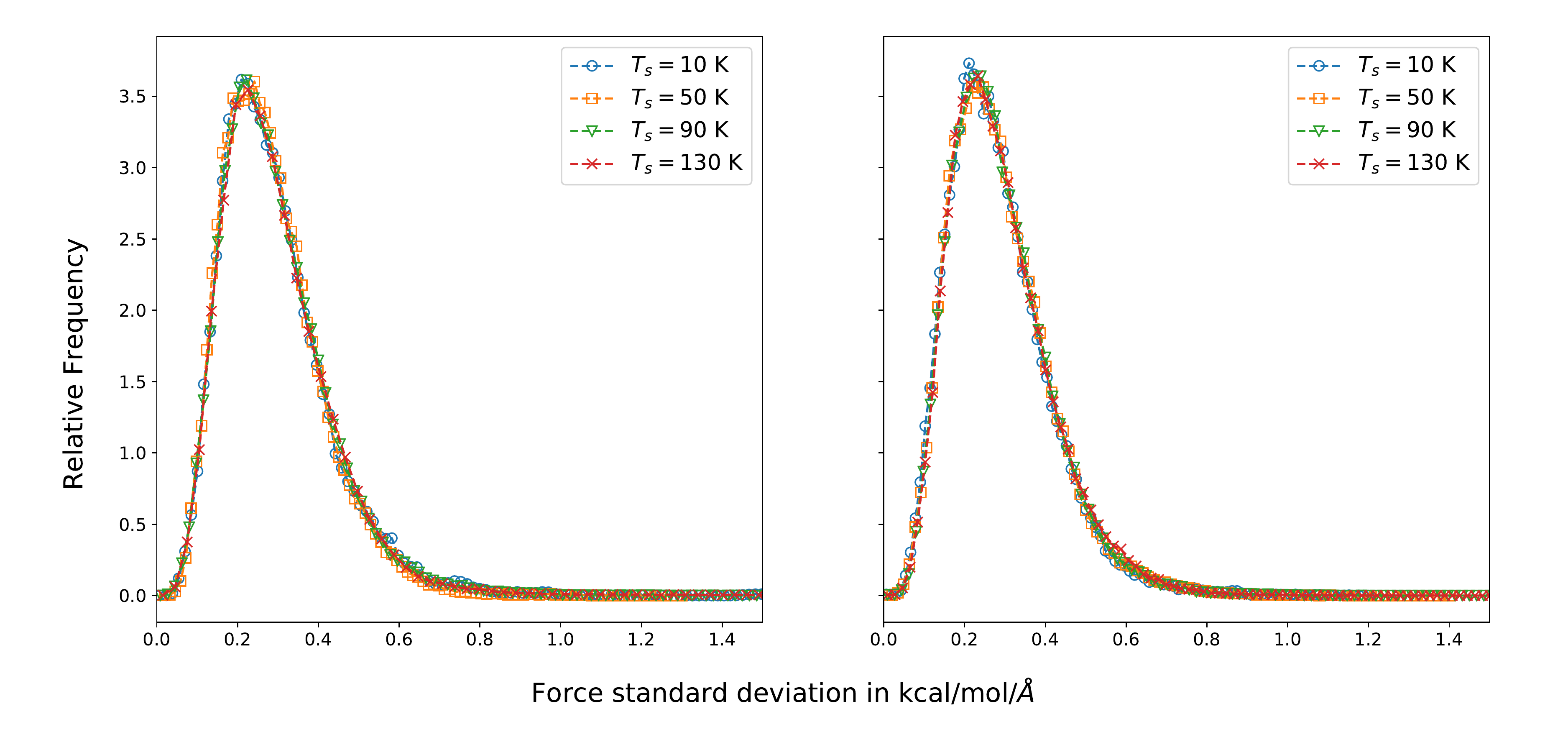}
    \caption{Distribution of standard deviation for atomic forces predicted by the committee for (\textit{left}) the high incoming energy of $14.47$ kJ/mol of adsorbate atom and (\textit{right}) the low incoming energy of $1.45$ kJ/mol of adsorbate atom. Distributions are given for four surface temperatures. The amount of atomic forces predicted with a deviation larger than $1$~kcal/mol/{\AA} is less than $0.44$~\%.}
    \label{fig:std_distribution_forces}
\end{figure*}

\begin{figure*}
    \centering
    \includegraphics[width=\linewidth]{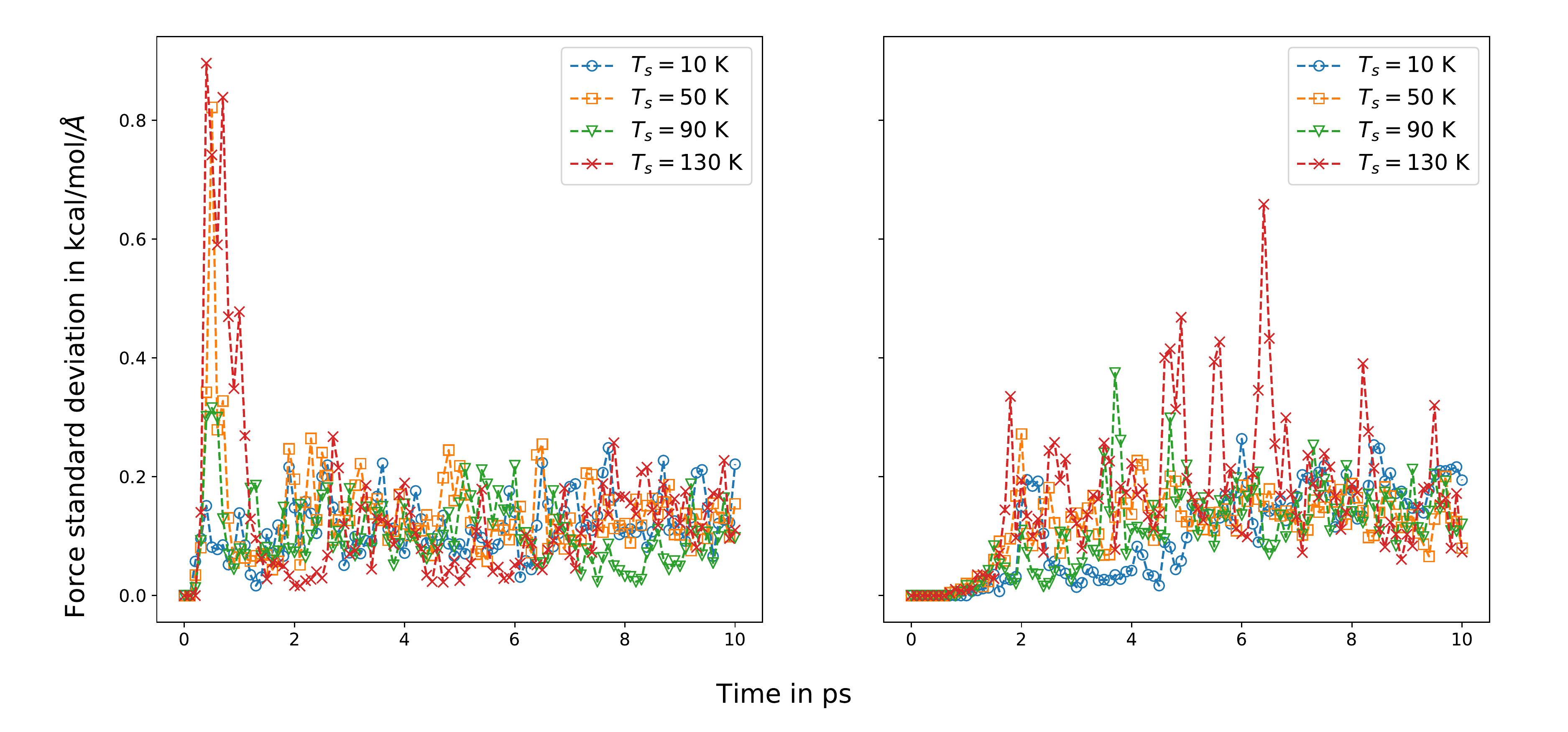}
    \caption{Standard deviation of atomic forces predicted for the adsorbate atom by the committee for (\textit{left}) the high incoming energy of $14.47$ kJ/mol of adsorbate atom and (\textit{right}) the low incoming energy of $1.45$ kJ/mol of adsorbate atom. Deviations are given for four surface temperatures.}
    \label{fig:std_adsorbate}
\end{figure*}

\begin{figure*}
    \centering
    \begin{minipage}{0.45\linewidth}
        \centering
        \includegraphics[width=\linewidth]{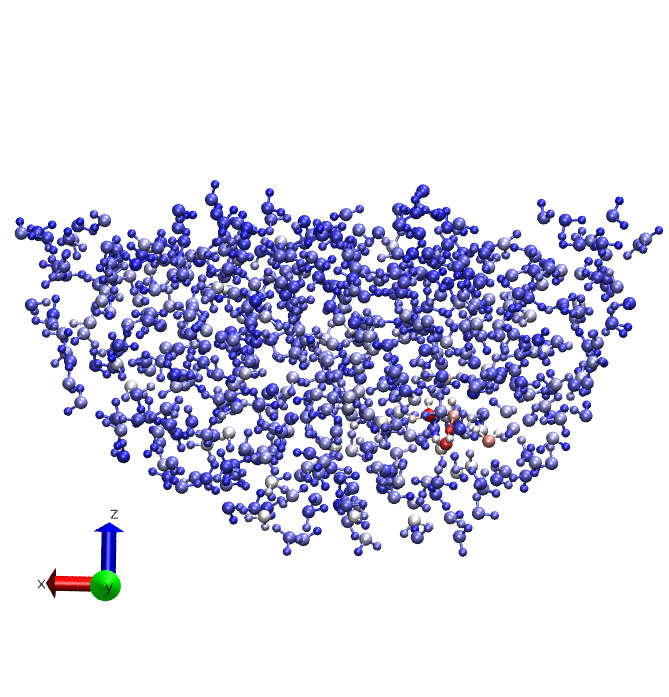}
    \end{minipage}\hfill
    \begin{minipage}{0.45\linewidth}
        \centering
        \includegraphics[width=\linewidth]{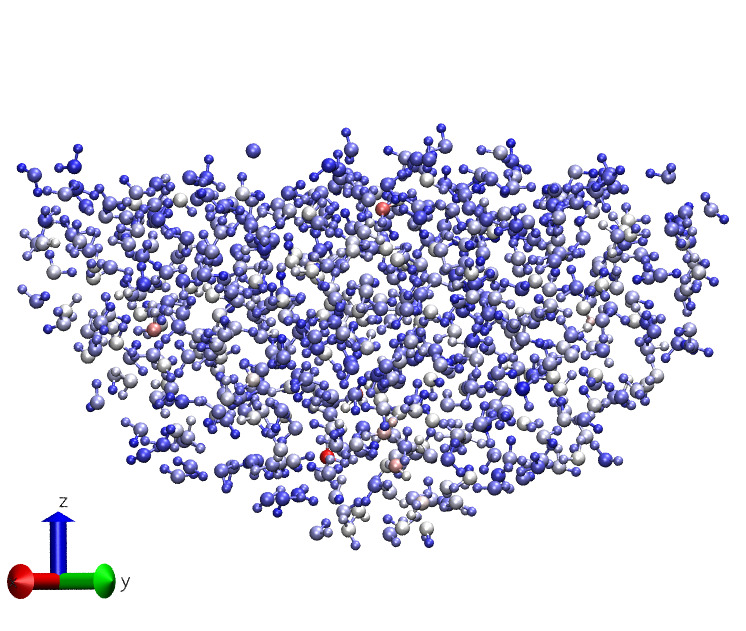}
    \end{minipage}
    \caption{\textit{Left:} Deviations of the predicted atomic forces for the surface temperature of $130$~K.  Blue color represents small errors starting at $0.04$~kcal/mol/{\AA}. Red color represents higher errors with the maximal value of $6.94$~kcal/mol/{\AA}. Gray color represents intermediate values of about $1$~kcal/mol/{\AA}. \textit{Right:} Deviations of the predicted atomic forces for the surface temperature of $50$~K.  Blue color represents small errors starting at $0.05$~kcal/mol/{\AA}. Red color represents higher errors with the maximal value of $1.08$~kcal/mol/{\AA}. Gray color represents intermediate values of about $0.5$~kcal/mol/{\AA}.}
    \label{fig:asw_std_forces}
\end{figure*}

To track the extrapolation three different models were trained on the same labeled training data. During the collision dynamics, every 100~fs all models were used to predict the energy and atomic forces of the respective structure. These were stored and analysed in the post-processing step. As a measure of disagreement between models we have chosen the standard deviation between atomic forces, which was defined as
\begin{equation}
    \sigma = \sqrt{\frac{1}{3 N_\text{models}} \sum_{i=1}^{N_\text{models}}\sum_{j \in \{x, y, z\}} \left( F_{ij} - \bar{F}_{j}\right)^2 },
\end{equation}
where $N_\text{models}$ is the number of models used to build the committee and $\bar{F}_j$ is the mean of the force components over the committee.
\figref{fig:std_distribution_forces} shows an example of the distribution of the standard deviation of atomic forces during the collision dynamics for four surface temperatures. In \figref{fig:std_distribution_forces} two different incoming energies are represented, 14.47~kJ/mol and 1.45~kJ/mol. Averaging over all collision trajectories we obtained a mean value for the force deviation of $0.31$~kcal/mol/{\AA} and its standard deviation equals to $0.27$~kcal/mol/{\AA}, which perfectly match with the presented distributions. Note that only about $0.44~\%$ of all predicted atomic forces have a deviation larger than 1~kcal/mol/{\AA}. To demonstrate how accurate the employed MLPs are, we also show the force deviations of the adsorbate atom during the whole MD simulation in \figref{fig:std_adsorbate}. In the figure four different surface temperatures and two different incoming energies, a high one and a low one, are represented. The high incomming energy along with a high surface temperature is close to a worst-case scenario concerning the accuracy. Here, the adsorbate enters the strongly repulsive region when two atoms approach each other, experiencing strong forces. Even in these extreme cases, the deviations are still lower than the desired accuracy with respect to the underlying DFT method, i.e. 1~kcal/mol/\AA{}. 

Values of $\sigma$, which are noticeably larger than the rest are indications of extrapolation, of a situation where the geometry is not covered sufficiently by the training set. The largest standard deviations are present in the bulk water atoms. To demonstrate this, we plot the system with atoms colored according to $\sigma$, see \figref{fig:asw_std_forces}. It can be observed that only a small portion of atoms show  large deviations in predicted forces. These are  mainly are the bulk atoms and close to the frozen atoms. These atoms are hardly relevant for the adsorption dynamics.  Therefore, we neglected the respective deviations during collision dynamics. However, the observed deviations can be further improved by including geometries with high $\sigma$ values as additional data in the training set. 

\section{Derivation of Temperature-Dependent Sticking Coefficients} \label{sec:derivation}

We present here the derivation of an analytic expression for \Eqref{eq:integration}, i.e:

\begin{equation} \label{eq:integration_2}
    S_T = \frac{1}{(kT)^2}\int_0^\infty P\left(E\right)Ee^{-\frac{E}{kT}}\mathrm{d}E,
\end{equation}
$P(E)$ can be linearly interpolated for an interval between $E_{a}$ and $E_{b}$ contained within the sampled data according to:
\begin{equation}
    P(E_{a})= P_{a}, \qquad P(E_{b})=P_{b}
\end{equation}
and
\begin{equation}
    P(E)= A+BE,
\end{equation}
with:
\begin{equation}
    A=P_{a} - \dfrac{E_{a}}{E_{b}-E_{a}}(P_{b}-P_{a}),
\end{equation}
\begin{equation}
    B=\dfrac{P_{b}-P_{a}}{E_{b}-E_{a}}.
\end{equation}
\Eqref{eq:integration_2} can now be decomposed into a sum of considered intervals:
\begin{equation}
    S_{T}=\sum_{i}S_{i},
\end{equation}
where:
\begin{equation} 
    S_{i} = \frac{1}{(kT)^2}\int_{E_{a}}^{E_{b}} P\left(E\right)Ee^{-\frac{E}{kT}}\mathrm{d}E.
\end{equation}
Now, considering $\beta=-1/kT$ one has:
\begin{equation}
    S_{i}= \beta^2 \int_{E_{a}}^{E_{b}} (A+BE)  Ee^{\beta E}\mathrm{d}E,
\end{equation}
which results in 
\begin{multline} \label{eq:last_si}
    S_{i} = A \left[ e^{\beta E_{b}} \left( \beta E_{b} - 1\right) - e^{\beta E_{a}}\left( \beta E_{a} -1\right) \right] \\
    +B\left[ e^{\beta E_{b}} \left( E_{b}^{2} \beta -2 E_{b} + \dfrac{2}{\beta}\right) -e^{\beta E_{a}}\left(E^{2}_{a}\beta -2 E_{a} + \dfrac{2}{\beta}\right)\right]. 
\end{multline}
\Eqref{eq:last_si} can be used for evaluation of the integral. In the error propagation, we take the derivatives of $S_{i}$ with respect to $P(E)$. The terms within square brackets in \Eqref{eq:last_si} do not depend on $P(E)$ so we can write:
\begin{equation}
    S_{i}=AC_{A}+BC_{B}
\end{equation}
The total variance, $s_{t}^{2}$, of the measure, considering the number of sub-intervals as measuring points, $N$, can be calculated according to the error propagation:
\begin{equation}
    s_{t}^{2}=\dfrac{1}{\sqrt{N}}\sum_{i}\left(\dfrac{\partial S_{i}}{\partial P_{a}}\right)^{2} s_{a}^{2}+ \left(\dfrac{\partial S_{i}}{\partial P_{b}}\right)^{2} s_{b}^{2},
\end{equation}
with $s_{a}^{2}$ being the variance of $P_{a}$ and $s_{b}^{2}$ being the variance of $P_{b},$ 
\begin{equation}
    \dfrac{\partial S_{i}}{\partial P_{a}}= C_{A}\dfrac{\partial A}{\partial P_{a}}+C_{B}\dfrac{\partial B}{\partial P_{a}},
\end{equation}
\begin{equation}
    \dfrac{\partial A}{\partial P_{a}}=\dfrac{E_{b}}{E_{b}-E_{a}}, \\
    \dfrac{\partial A}{\partial P_{b}}=\dfrac{-E_{a}}{E_{b}-E_{a}},
\end{equation}
\begin{equation}
    \dfrac{\partial B}{\partial P_{a}}=\dfrac{-1}{E_{b}-E_{a}}, \\
    \dfrac{\partial B}{\partial P_{b}}=\dfrac{1}{E_{b}-E_{a}},
\end{equation}
\begin{equation}
    \dfrac{\partial S_{i}}{\partial P_{a}}=\dfrac{C_{A}E_{b}-C_{B}}{E_{b}-E_{a}}, \\
    \dfrac{\partial S_{i}}{\partial P_{b}}=\dfrac{-C_{A}E_{b}+C_{B}}{E_{b}-E_{a}}.
\end{equation}
Finally, we can approximate $s_{a}^{2}$ and $s_{b}^{2}$ from our confidence intervals in $P(E)$. Doing the approximation of treating $P(E)$ as a Gaussian we can determine that $s^{2}\sim (L/4)^{2}$ with $L$ being the width of the confidence interval.


\bsp	
\label{lastpage}
\end{document}